\documentclass[preprint,preprintnumbers,amsmath,floatfix]{revtex4}
\usepackage{amsfonts}    
\usepackage{amssymb}
\usepackage{latexsym}
\usepackage{eepic}
\usepackage{graphicx}
\usepackage{subfig}
\usepackage{subfig}
\usepackage{multirow}
\usepackage{rotating,booktabs} 
\usepackage{pdflscape}
\usepackage[active]{srcltx}
\usepackage[colorlinks=true]{hyperref}
\usepackage{caption}
\usepackage{color}
\usepackage{ulem}
\usepackage{soul}
\newcommand{\re}[1]{(\ref{#1})}

\newcommand{\al}{\alpha}

\newcommand{\ep}{\epsilon}
\newcommand{\Si}{\Sigma}

\newcommand{\si}{\sigma}

\newcommand{\de}{\delta}

\newcommand{\De}{\Delta}

\newcommand{\rar}{\rightarrow}

\newcommand{\non}{\nonumber}
\newcommand{\m}{\,\,}

\begin{document}

\title{{H$_2^+$, HeH and H$_2$}: approximating potential curves, calculating
rovibrational states}

\author{Horacio~Olivares-Pil\'on} \email{horop@xanum.uam.mx}
\affiliation{Departamento de F\'isica, Universidad Aut\'onoma
  Metropolitana-Iztapalapa, Apartado Postal 55-534, 09340 M\'exico,
  D.F., Mexico}

\author{Alexander~V.~Turbiner}
\email{turbiner@nucleares.unam.mx}
\affiliation{Instituto de Ciencias Nucleares, Universidad Nacional
Aut\'onoma de M\'exico, Apartado Postal 70-543, 04510 M\'exico,
D.F., Mexico{}}


\vskip 1cm

\begin{abstract}

Analytic consideration of the Bohr-Oppenheimer (BO) approximation for diatomic molecules is proposed:
accurate analytic interpolation for potential curve consistent with its rovibrational spectra is found.

It is shown that in the Bohr-Oppenheimer approximation for four lowest electronic states
$1s\si_g$ and $2p\si_u$, $2p \pi_u$ and $3d \pi_g$ of H$_2^+$, the ground
state X$^2\Si^+$ of HeH and the two lowest states $^1\Si^+_g$ and $^3\Si^+_u$ of H$_2$,
the potential curves can be analytically interpolated in full range of internuclear distances $R$
with not less than {4-5-6} figures.
Approximation based on matching the Taylor-type expansion at small $R$
and a combination of the multipole expansion with one-instanton type contribution at large distances
$R$ is given by two-point Pad\'e approximant.
The position of minimum, when exists, is predicted within 1$\%$ or better.

For the molecular ion H$_2^+$ in the Lagrange mesh method, the spectra of vibrational, rotational
and rovibrational states $(\nu,L)$ associated with $1s\si_g$ and $2p\si_u$, $2p \pi_u$ and $3d \pi_g$
potential curves is calculated. In general, it coincides with spectra found via
numerical solution of the Schr\"odinger equation (when available) within six figures.
It is shown that $1s\si_g$ curve contains 19 vibrational states $(\nu,0)$, while $2p\si_u$ curve
contains a single one $(0,0)$ and $2p\pi_u$ state contains 12 vibrational states $(\nu,0)$.
In general, $1s\si_g$ electronic curve contains 420 rovibrational states,
which increases up to 423 when we are beyond BO approximation.
For the state $2p\si_u$ the total number of rovibrational states (all with $\nu=0$) is equal
to 3, within or beyond Bohr-Oppenheimer approximation. As for the state $2p\pi_u$ within the
Bohr-Oppenheimer approximation the total number of the rovibrational bound states is equal to 284.
The state $3d\pi_g$ is repulsive, no rovibrational state is found.

It is confirmed in Lagrange mesh formalism the statement that the ground state potential curve
of the heteronuclear molecule HeH does not support rovibrational states.

Accurate analytical expression for the potential curves of the hydrogen molecule
H$_2$ for the states $^1\Si^+_g$ and $^3\Si^+_u$ is presented. The ground state
$^1\Si^+_g$ contains 15 vibrational states $(\nu,0), \nu = 0-14$. In general, this state
supports 301 rovibrational states. The potential curve of the state $^3\Si^+_u$ has a shallow
minimum: it does not support any rovibrational state, it is repulsive.

\end{abstract}


\maketitle

\centerline{\bf INTRODUCTION}

\vskip 1cm

A chance to "integrate out" effectively the electronic degrees of freedom is a remarkable
feature of the Bohr-Oppenheimer approximation. The original problem say for simplicity
of diatomic molecule is reduced to two-body problem with an effective potential called
{\it the (electronic) potential curve} $V(R)$ with the Hamiltonian (after the center-of-mass
separation) of the form
\begin{equation}
\label{HBO}
  {\cal H}(R)\ =\ \frac{P^2}{2} + V(R) + \frac{L(L+1)}{R^2}\ ,
\end{equation}
which describes nuclear motion, here $L$ is angular momentum, the reduced mass for simplicity is placed
equal to one and the momentum $P=-i \nabla_R$, $\hbar=1$. The (electronic) potential curve $V(R)$ depends on the original state of the diatomic molecule and usually is known numerically only.
For a long time it was a challenge to find out how to interpolate analytically a potential curve with reasonably high accuracy in full range of internuclear distances. In the present paper we consider
three simplest diatomic molecular systems, two are homonuclear H$_2^+$ and H$_2$ and
another one is heteronuclear, HeH, and construct simplest the analytical approximations
of the potential curves.

The H$_2^+$ molecular ion is the simplest molecular system which exists in Nature. It plays
a fundamental role in different physics sciences:
atomic-molecular physics, in laser and plasma physics being also a traditional example
of two-center Coulomb system of two heavy Coulomb charges $Z$ and electron, $(Z, Z, e)$ in Quantum Mechanics (see e.g. \cite{LL}). It represents also the simplest diatomic molecule.
Due to the fact that the proton is much heavier than electron the problem is
usually explored in the static approximation - the Bohr-Oppenheimer approximation
of the zero order - where the protons are assumed to be infinitely heavy.
Contemporary theory of low-lying states of H$_2^+$ in the Bohr-Oppenheimer approximation
(electronic structure and radiative transitions) is presented at \cite{OT:2016}.
It is based on highly accurate locally approximation of the lowest eigenfunctions
proposed in \cite{Turbiner:2011}. However, potential curves of the electronic states remained
known numerically.

In turn, (HeH) (and its ions) represents the simplest neutral heteroatomic molecule(s).
It was intensely studied numerically, see e.g. \cite{MWB:2002} and references therein,
where a shallow van der Waals minimum was found. In general, it is as unstable and the
ground state potential curve is repulsive.

As for two-electron molecules, the simplest (neutral) one is the hydrogen molecule H$_2$, which plays
extremely important role in Nature. In particular, its importance relies on the fact that H$_2$ is presented significantly in planetary atmospheres, in the Earth one, for instance. The ground state $^1\Si^+_g$ displays
a well-pronounced minimum at finite internuclear distance, while the first excited state
$^3\Si^+_u$ develops a shallow minimum at most, which does not support rovibrational states
being repulsive.

Present paper is aimed to construct the simplest possible analytic approximations of potential curves
of the low-lying electronic states of H$_2^+$, (HeH) and H$_2$ in full range of internuclear distances,
which reproduce not less than 4-5-6 figures in numerically-found potential curves.
The same time we require that the rovibrational spectra associated with given potential curve is reproduced
with not less than 5 figures, thus, being well inside of the domain of applicability of (static) Bohr-Oppenheimer approximation.

Specifically, based on analytical approximations we develop the theory of vibrational and rotational
states for four low-lying electronic states for H$_2^+$, for the ground state potential curve of (HeH) and
two lowest ones of H$_2$.

Atomic units are used throughout, in particular, for distances, although the energy is given in Rydbergs.


\section{Generalities: THE MOLECULAR ION H$_2^+$}

The Schr\"odinger equation, which describes the electron in the field of two fixed centers of the
charges $Z_1, Z_2$ at the distance $R$, is of the form
\begin{equation}
\label{Sch}
    \left(-\De - \frac{2 Z_1}{r_1}- \frac{2 Z_2}{r_2}\right)\Psi \ =\ E' \Psi\ ,\
    \Psi \in L^2 ({\bf R^3})\ ,
\end{equation}
where $E'=(E - \frac{2 Z_1 Z_2 }{R})$, $E(R)$ is the total energy, both $e', E(R)$ are in Rydbergs, $r_{1,2}$ are the distances from electron to first (second) center, respectively. From physical point of view, we study the motion of electron in the field of two Coulomb wells situated on the distance $R$. If $Z_1=Z_2$ the wells become identical - any eigenstate is characterized by a definite parity with respect to permutation of wells (centers). Furthermore, at $R \rar \infty$, when the barrier gets large and tunneling becomes exponentially-small, the phenomenon of pairing should occur: the spectra of positive parity states is almost degenerate with the spectra of negative parity states, asymptotically both are equal to the energy spectra of hydrogen atom $E_H$. For each pair the energy gap should be
exponentially-small,~$\sim e^{-a R}$ where $a$ is a parameter. Dissociation energy is given by
\begin{equation}
\label{Ediss}
    \tilde E\ =\ E(R)\ -\ E_H\ .
\end{equation}

We focus on the case of unit charges $Z_1=Z_2=1$ - the only case where bound states occur -
it corresponds to H${}^+_2$ molecular ion.



\section{H${}^+_2$: The lowest states potential curves}

\subsection{Energy gap between $1s\si_g$ and $2p\si_u$ states}

The Born-Oppenheimer approximation leads to the concept of electronic potential
curve, which has the meaning of the total energy of the system H${}^+_2$
at fixed internuclear distance $R$. Thus, the problem to
find a potential curve is reduced to finding spectra of electronic
Schr\"odinger equation (\ref{Sch}), where $R$ plays a role of
parameter. Since the potential in (\ref{Sch}) is a double-well
potential with degenerate minima, it is natural to study the energy
gap, which is the distance between two lowest eigenstates,
\begin{equation}
\label{gap}
      \De E\ =\ E_{2p\si_u}\ -\ E_{1s\si_g}\ .
\end{equation}
The goal of the Sections II.A-B is to refine the results obtained in \cite{OT:2016}.

For small $R$ it was found long ago the expansion with finite radius of convergence ~\cite{BB:1965,BS:1966,K:1983}
\footnote{Due to controversy in literature we use in \cite{OT:2016} a slightly incorrect coefficient (in $\sim 1 \% $) in front of $R^2$ term:\ $\frac{27}{5}$. It did not really lead to a change of conclusions in \cite{OT:2016}}
\begin{equation}
\label{DE-0}
  \De E\ =\ 3\ -\ \frac{82}{15}\,R^2\ +\frac{32}{3}\,R^3+\ O\,(R^4)\ ,
\end{equation}
while at large $R$ the corresponding expansion, see \cite{LL}, Ch.XI and \cite{OV:1964,DP:1968,Cizek:1986},
\begin{equation}
\label{DE-infty}
  \De E\ =\ \frac{8}{e}\, R\, e^{-R}\left(1 + \frac{1}{2R} -\frac{25}{8\, R^2} + \cdots\right)\ + O(e^{-2R}).
\end{equation}
It looks like the multi-instanton expansion where $R$ plays a role of the classical
action. Seemingly, the series in pre-factor is asymptotic - it has zero radius of convergence in $1/R$.

\begin{figure}[!thb]
\includegraphics[scale=1.4]{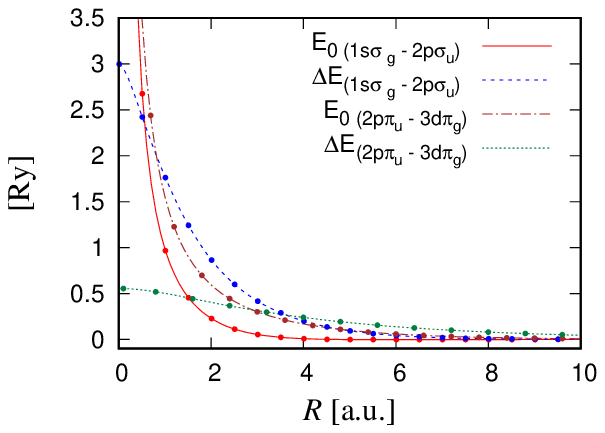}
\caption{H$_2^+$:\ $E_0$ and $\De E$ between the states  $1s\si_g$ - $2p\si_u$
and $2p\pi_u$ - $3d\pi_g$.  Calculated energies are marked by dots and the solid curves
are the fits~\re{eRfit} (solid line), \re{f2} (dashed line), \re{e0R2p1} (dot-dash line)
and  \re{f22} (dotted line).}
\label{e0anddE}
\end{figure}

Now we take data for potential curves of the $1s\si_g$ and $2p\si_u$
states, see Tables 1,2 \cite{OT:2016}, calculate the difference $\De E$ and interpolate between
small and large distances using the Pad\'e type approximation $e^{-R-1}\ \mbox{Pade}[N+1/N](R)$,
where $\mbox{Pade}[N+1/N](R)$ is meromorphic function, $p_{N+1}(R)/q_{N}(R)$. In general,
$\De E$ is smooth, slow-changing curve with $R$, exponentially-vanishing at $R \rar \infty$, see below
Fig.~\ref{e0anddE}.

Taking several different values $N$ we found that $N=7$ is the smallest which provides the quality of fit
we would like to have, see below,
\[
 \De E\ =\ e^{-R-1}\ \mbox{Pade}[8/7](R)\ ,
\]
or explicitly,
\begin{equation}
\label{f2}
      \De E \ =\ e^{-R-1}\ \frac{3e+a_1 R+a_2 R^2+a_3 R^3+a_4 R^4+a_5 R^5+a_6 R^6
      +a_7 R^7+8 R^8} {1+\al_1 R+\al_2 R^2+b_3 R^3+b_4 R^4+b_5 R^5+\al_3 R^6+ R^7}\ ,
\end{equation}
where three constraints
\begin{equation}
\begin{array}{l}
   \mbox{$\alpha_1= (a_1-3e)/(3e)$}\ ,\\
   \mbox{$\alpha_2= (-a_1+a_2 +\frac{209e}{30})/(3e)$}\ ,\\
   \mbox{$\alpha_3=(a_7-4)/8$}\ ,
\end{array}
\end{equation}
are imposed, which guarantee that the appropriate expansions of (\ref{f2}) reproduce correctly the $R^0$, $R^1$ and $R^2$ terms in (\ref{DE-0}) and the two terms in (\ref{DE-infty}). Eventually, (\ref{f2}) has ten free parameters which are fixed by making fit of numerical data with~(\ref{f2}) with minimal $\chi^2$. As the result those ten free parameters take values:
\begin{equation}
\label{parmsfits}
\begin{array}{ll}
a_1 = 446.5741\ ,\  &  a_6 =  214.0609\ , \\
a_2 = 905.1538\ ,\  &  a_7 = -52.89581\ , \\
a_3 = 223.2718\ ,\  &  b_3 =  38.57209\ , \\
a_4 = 307.0596\ ,\  &  b_4 = -38.28025\ , \\
a_5 =-235.4166\ ,\  &  b_5 =  31.65441\ ,
\end{array}
\end{equation}
where the seven shown figures are significant. Note that the parameters (\ref{parmsfits}) are much smaller than ones find in \cite{OT:2016}.
This fit gives, in general, 6-7\,figures at the whole range $R \in [0,40]$~a.u., see Table \ref{tede}, and furthermore, up to 9\,d.d. for large $R \in [20,40]$\,a.u.
\ (see for illustration Table~\ref{tede} and Fig.~\ref{e0anddE}, blue curve). Saying differently,
the largest absolute difference between exact and fitted energy gaps occurs at 5th decimal digit in domain
$R \in [0.5, 9.0]$~a.u. It gets even more accurate outside of this domain.

\begin{center}
\begin{table}
\caption{H$_2^+$:\ $E_0$ (\ref{E0-sum}) and $\De E$ (\ref{gap}) between the ground $1s\si_g$ and first excited $2p\si_u$  in Ry as a function of the internuclear distance $R$ compared to the results of the fits~\re{eRfit}~and~\re{f2}.}
\label{tede}
\begin{tabular}{r| ll | ll}
\hline\hline
  &\multicolumn{2}{c}{$E_0=\frac{1}{2}(E_{2p\si_u}+E_{1s\si_g})+1$}\vline& \multicolumn{2}{c}{$\Delta E=E_{2p\si_u}-E_{1s\si_g}$}\\ \cline{2-5}
  R& \qquad data &Fit~\re{eRfit}& \qquad data &Fit~\re{f2}\\
\hline
0.1 &\ 18.5210904842   &18.521096     &\ 2.9551492517 &\ 2.955137 \\
0.5 &\ \m2.7481265348  &\m2.748074    &\ 2.4362050695 &\ 2.436192 \\
1.0 &\ \m0.9834000615  &\m0.983519    &\ 1.7739453766 &\ 1.773930 \\
2.0 &\ \m0.2298313933  &\m0.229534    &\ 0.8701996446 &\ 0.870186 \\
3.0 &\ \m0.0543521359  &\m0.054627    &\ 0.4189557280 &\ 0.418955 \\
4.0 &\ \m0.0083644769  &\m0.008011    &\ 0.2010684887 &\ 0.201072 \\
5.0 &\ -0.0017119084   &-0.001535	    &\ 0.0942573639 &\ 0.094260 \\
6.0 &\ -0.0026129408   &-0.002134	    &\ 0.0426503122 &\ 0.042641 \\
7.0 &\ -0.0018657168   &-0.001514	    &\ 0.0186445834 &\ 0.018645 \\
8.0 &\ -0.0011764042   &-0.000989	    &\ 0.0079287460 &\ 0.007935 \\
9.0 &\ -0.0007392824   &-0.000649	    &\ 0.0033032469 &\ 0.003309 \\
10.0&\ -0.0004797975   &-0.000437	    &\ 0.0013553207 &\ 0.001359 \\
30.0&\ -5.581483$\times10^{-6}$ &-5.569$\times10^{-6}$ & $\quad < 10^{-10}$   &\quad $ < 10^{-10}$  \\
%
\hline\hline
\end{tabular}
\end{table}
\end{center}

\subsection{The ground state $1s\si_g$ and the first excited state $2p\si_u$}

For the lowest state $1s\si_g$, the behavior of the potential curve
$E_{1s\si_g}$ at the two asymptotic limits of small and large distances is
well known.  For $R\rightarrow 0$ the dissociation energy is given
by~\cite{BB:1965,BS:1966,K:1983}
\begin{equation}
\label{R0-0}
    {\tilde E}_{1s\si_g}^{(0)}\ =\ \frac{2}{R}\ -\ 3\ +\ \frac{16}{3}\,R^2\ -\ \frac{32}{3}\,R^3\ +\ O(R^4 \log R) \ ,
\end{equation}
it can be found in perturbation theory in powers of $R$. Note the linear in $R$ term is absent. In turn,
for the dissociation energy at $R\rightarrow \infty$ the expansion reads~\cite{OV:1964,DP:1968,Cizek:1986}
\begin{eqnarray}
\label{vinfg}
  {\tilde E}_{1s\si_g}^{(\infty)} & = & -\frac{9}{2\, R^4}\ -\ \frac{15}{R^6}\ -\ \frac{213}{2\, R^7}\ +
  \cdots\\\non
  && - 4R\,e^{-R-1}\left[1+\frac{1}{2\, R}-\frac{25}{8\, R^2}-\frac{131}{48\, R^3}
  -\frac{3923}{384\, R^4}\ +\ \cdots\right]\ +\ O(e^{-2 R})\ ,
\end{eqnarray}
where the first sum represents the multipole expansion, the second one is a type of one-instanton contribution etc. Note that in the multipole expansion the term $R^{-5}$ (next-after-leading) is absent.

As for the lowest state of the negative parity $2p\si_u$ large and small $R$-distance expansions
are known as well,
\begin{equation}
\label{R0-1}
    {\tilde E}_{2p\si_u}^{(0)}\ =\ \frac{2}{R}\ -\ \frac{2}{15}R^2\ +\ \ldots\ ,
\end{equation}
at $R \rar 0$, where the next-after-leading terms, $O(1)$ and $O(R)$, are absent,
see~\cite{B:1958}.
The behavior for $R \rar \infty$ is the same as one given by Eq.~\re{vinfg} with sign changed from minus to plus in front of the exponentially-small term $\sim e^{-R}$.

Let us consider the sum of potential curves for
$1s\si_g$ and $2p\si_u$ states,
\begin{equation}
\label{E0-sum}
     E_0\ \equiv \ \frac{{\tilde E}_{1s\si_g}+{\tilde E}_{2p\si_u}}{2}\ .
\end{equation}
Its corresponding expansions are
\begin{equation}
\label{R0-sum}
 E_0\ =\ \frac{2}{R}\ -\ \frac{3}{2}\ +\ \frac{13}{5}R^2\ +\ \ldots\ ,
\end{equation}
at $R \rar 0$, see \cite{B:1958}, where the linear in $R$ term is absent, and
\begin{equation}
\label{Rinfty-sum}
 E_0\ =\ -\frac{9}{2\, R^4}\ -\ \frac{15}{R^6}\ -\ \frac{213}{2\, R^7}\ +\ \ldots\
 +\ O(e^{-2R})\ ,
\end{equation}
at $R \rar \infty$, where again the term $R^{-5}$ is absent. The first expansion (\ref{R0-sum})
has a finite radius of convergence, see e.g. \cite{K:1983}, while the second one (\ref{Rinfty-sum}) (the half-sum of two multipole expansions) has zero radius of convergence.

Now we assume that two-instanton contribution, $\sim e^{-2R}$ at large $R$ (and possible higher exponentially-small contributions), can be neglected and construct the analytic approximation
for $E_0$ which mimics the two asymptotic limits (\ref{R0-sum}), (\ref{Rinfty-sum}). The most convenient
way to do it is to use Pad\'e type approximation (ratio of two polynomials)
$E_0(R)=\frac{1}{R}\ \mbox{Pade}[N/N+3](R)$ with a certain $N$.
Concrete fit was made for $N=5$, where the Pad\'e approximation is of the explicit form
$E_0(R)=\frac{1}{R}\ \mbox{Pade}[5/8](R)$,
\begin{equation}
\label{eRfit}
   E_0 =\frac{2+ a_1R+a_2R^2 +a_3R^3 +a_4R^4 -9R^5}{R (1 +\al_1R
  +\al_2R^2+b_3R^3+b_4R^4+b_5R^5-\al_3R^6-\al_4R^7+2R^8)}\ ,
\end{equation}
with four constraints imposed,
\begin{equation}
\begin{array}{l}
   \mbox{$\al_1= (a_1+3/2)/2$}\ ,\\
   \mbox{$\al_2= (6a_1+8a_2+9)/16$}\ ,\\
   \mbox{$\al_3= 2(a_3+30)/9$}\ ,\\
   \mbox{$\al_5= 2a_4/9$}\ ,
\end{array}
\end{equation}
see below, thus, (\ref{eRfit}) depends eventually on seven free parameters. Above constraints guarantee that the three coefficients in front of $R^{-1}$, $R^{0}$ and $R^{1}$ of expansion at $R\rightarrow
0$, see \re{R0-sum} and the three coefficients in front of $R^{-4}$, $R^{-5}$ and $R^{-6}$ at
$R\rightarrow \infty$, in the $1/R$-expansion (\ref{Rinfty-sum}) are all exact.
After making the fit with~(\ref{eRfit}),  we arrive to concrete values of these seven free parameters:
\begin{equation}
\label{parmsfits1}
\begin{array}{ll}
a_1 =  267.095\ ,\  & b_3 =  124.971\ ,\\
a_2 =  375.335\ ,\  & b_4 =  -57.4398\ ,\\
a_3 = -180.965\ ,\  & b_5 = 10.5173\ ,\\
a_4 =  60.700\ ,\  & \\
\end{array}
\end{equation}
c.f. (47) in \cite{OT:2016}.
It provides not less than \,5-6\,figures in $E_0$ for whole studied domain $R \in [1,40]$~a.u.,
see Table~\ref{tede}  and Fig.~\ref{e0anddE}. Saying differently, the largest absolute difference between exact and fitted energies occurs at 5th decimal digit in domain $R \in [0.5, 9.0]$~a.u. It gets smaller outside of this domain.

\begin{figure}[!thb]
\includegraphics[scale=1.5]{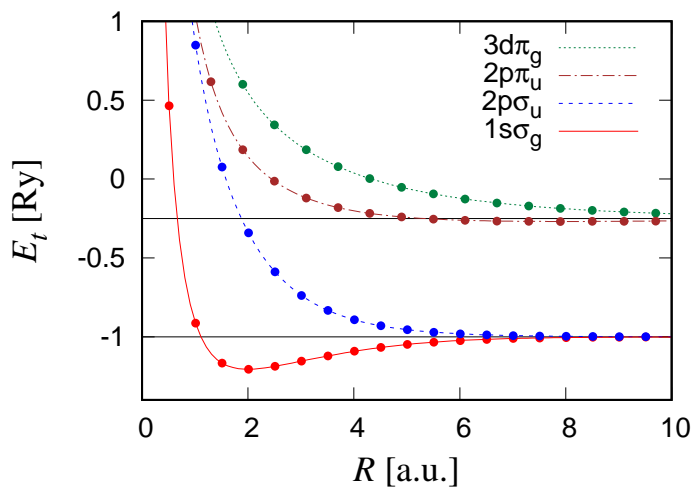}
\caption{H$_2^+$:\ Potential energy curves. Points are
         the calculated values and curves are fits: $1s\si_g$ \re{pt1s} (solid line),
         $2p\si_u$ \re{pt2p} (dashed line), $2p\pi_u$ \re{f2ppu} (dot-dash line) and
         $3d\pi_g$ \re{f3dpg} (dotted line).}
\label{pch2pRy}
\end{figure}

The potential curve for the ground state $1s\si_g$ can be constructed from \re{eRfit} and \re{f2} by taking
\begin{equation}
\label{pt1s}
   E_{1s\si_g}\ =\ E_0 - \frac{1}{2}\De E\ .
\end{equation}
This expression reproduces 4-5-6 figures in energy for the whole domain $R \in [0, 40]$~a.u., when comparing with the exact values, see Table 1 \cite{OT:2016}, for illustration see
Fig.~\ref{pch2pRy}. It differs in 5-6 figure. It is quite remarkable that the minimum of the
potential curve is predicted at $E^{(fit)}_{min} = -1.205\,56$~Ry (c.f. $E_{min}^{exact}=-1.205\,27$~Ry)
while its location is  $R^{(fit)}_{eq}=1.996\,84$~a.u. (cf. $R_{eq}^{exact}= 1.997\,19$~a.u.).

The asymptotic expansions of Eq.~(\ref{pt1s})
are given by
\begin{eqnarray}
      &{\tilde E}_0^{}&=\frac{2}{R}\ -\ 3\ +\ 4.413\,R^2\ +\cdots \ ,\\
      &{\tilde E}_{\infty}&=-\frac{9}{2R^4}-\frac{15}{R^6}+\frac{110.166}{R^7}+\cdots
      -4Re^{-R-1}\left[ 1 + \frac{1}{2\,R}-\frac{1.3408}{R^2}\cdots\right] \ ,
\end{eqnarray}
which are in complete agreement with the first three terms at
$R\rightarrow 0$, and with the first three terms in the $1/R$
expansion and two terms in $1/R$ expansion of the pre-factor to
$e^{-R}$ for $R\rightarrow \infty$ (cf.~\re{R0-0} and \re{vinfg}).

Similarly, the potential curve for the excited state $2p\si_u$ is
restored from \re{eRfit} and \re{f2} by taking
\begin{equation}
\label{pt2p}
   E_{2p\sigma_u}\ =\ E_0 + \frac{1}{2}\De E\ .
\end{equation}
This expression also reproduces 4-5-6 figures when comparing with the exact energy,
see Table 2 in \cite{OT:2016} and for illustration Fig.~\ref{pch2pRy}.
The asymptotic expansions of Eq.~(\ref{pt2p})
are given by
\begin{eqnarray}
     &{\tilde E}_0^{}&=\frac{2}{R}\ -1.0536\,R^2\ +\cdots \ ,\\
     &{\tilde E}_{\infty}&=-\frac{9}{2R^4}-\frac{15}{R^6}+\frac{110.166}{R^7}+\cdots
      +4Re^{-R-1}\left[ 1 + \frac{1}{2\,R}-\frac{1.3408}{R^2}\cdots\right] \ ,
\end{eqnarray}
which are in complete agreement with the first three terms at
$R\rightarrow 0$ (cf.~\re{R0-0}), and three terms in $1/R$ expansion
and two terms in $1/R$ expansion of the pre-factor to $e^{-R}$ for
$R\rightarrow \infty$ (cf.~\re{vinfg}).

\subsection{Rovibrational states associated with the ground state curve $1s\si_g$}

Inside of the Born-Oppenheimer approximation the knowledge of potential electronic curve for $1s\si_g$ allows us to find the vibrational and rotational as well as rovibrational states by solving the equation
\begin{equation}
\label{vrs}
\left[-\frac{1}{\mu}\frac{d^2}{dR^2} +\frac{L(L+1)}{\mu R^2}+ V(R)\right]\phi(R)= E_{\nu L}\,\phi(R)\,,
\end{equation}
where $\mu=M_p/2$ is the reduced mass of the two protons, $V(R)$ is the total electronic energy for $1s\si_g$, $\nu$ and $L$ are the vibrational and rotational quantum numbers, respectively. In particular, the analytic approximation for the potential curve $V(R)=E_{1s\si_g}(R)$~\re{pt1s} made out the expressions \re{eRfit} and \re{f2} allows us to calculate the rovibrational energies $E_{\nu L}$
associated with the bound state $1s\si_g$ for different values of $\nu$ and $L$. It is done by solving
the one-dimensional differential equation~\re{vrs} using the Lagrange-mesh method, see e.g.~\cite{Baye:2015}. Then the spectra can be compared with a numerical solution of the Schr\"odinger equation (\ref{vrs})
with numerically-found potential $V(R)$. Simple estimate shows that the energy spectra obtained in these two methods should differ in 6th figure (or even beyond).

Table~\ref{t1ssgv} presents some of the studied vibrational states for $L=0$ until $L=35$ in increasing steps of $5$; the proton reduced mass $\mu=918.048$ is taken.
For $L=0$, second column displays the results given by Beckel~{\it et al.}~\cite{BHP:1970}. In general,
the agreement when comparing with our results is within $10^{-3}$. In all cases the first column, second lines show the rovibrational energies calculated by Moss~\cite{M:1993}, but where the finite mass effects are taken into account, thus, being beyond BO approximation.
It must be noted that these finite mass effects due to finiteness of the proton and electron masses usually change the fifth figure in the energy. It indicates the real accuracy of the Born-Oppenheimer approximation, which is four figures. Hence, it seems physically {\it irrelevant} to find potential curves with more than five figures (as it was done in the past, in particular, in \cite{BHP:1970}), when we are in static approximation. Note that beyond static approximation the potential curves do not exist(!).

In general, our approximation in the most cases agrees with old results by Beckel~{\it et al.}~\cite{BHP:1970} for $L=0$ within four-five figures - it provides relevant description of spectra of vibrational states H$_2^+$ within applicability of non-relativistic quantum mechanics in the Born-Oppenheimer approximation. Note our spectra also agrees with results by Moss~\cite{M:1993} within five figures!

Note that the existence of 20th vibrational state predicted by Moss with binding energy $\sim 6 \times 10^{-6}$\,a.u. goes beyond the applicability of Born-Oppenheimer approximation. Thus, the result obtained in \cite{BHP:1970} can not be trusted. We should conclude that within the Born-Oppenheimer approximation
the potential curve for $1s\si_g$ state keeps 19 vibrational states. Taking into account the finite mass corrections the number of vibrational states increases to 20.

For $L>0$ the agreement between our results in BO approximation and those by Moss \cite{M:1993} reduces to four and even (sometimes) to three figures, see Table \ref{t1ssgv}. Maximum angular momentum $L$, which still keeps vibrational bound state, is equal to $L_{max}=35$. It keeps a single vibrational state with $\nu=0$ with dissociation energy 0.00541\,Ry. For $L>35$ rovibrational bound states occur neither in BO approximation nor beyond
\footnote{Note for completeness that beyond the Born-Oppenheimer approximation
Moss [14] reported 14 quasi-bound rovibrational states for $L > 35$ and 44 quasi-bound rovibrational states for $L \leq 35$. The questions related to the existence of quasi-bound states are beyond the scope of the present paper}.

\begin{table}
 \caption{Rovibrational states $E_{\nu,L}$ of the ground electronic state $1s\si_g$ of the
 molecular ion H$_2^+$, all energies are in Ry. For $E_{\nu0}$ (the first two columns) comparison of our results
 (first line, first column) is done with
 Moss~\cite{M:1993} (non-adiabatic, rounded) (second line, first column)
 and with Beckel {\it et al.}~\cite{BHP:1970}, (adiabatic, second line, second column).
 For $L=5, \ldots 35$ our results (first lines) are compared with \cite{M:1993} (second lines)
 }
\label{t1ssgv}
{\scriptsize
\begin{center}
\begin{tabular}{rlllllllll}
\hline\hline
$\nu$&$E_{\nu0}$[Ry]&\cite{BHP:1970}&$E_{\nu,5}$[Ry]&$E_{\nu,10}$[Ry]&$E_{\nu,15}$[Ry]&$E_{\nu,20}$[Ry]&$E_{\nu,25}$[Ry]&$E_{\nu,30}$[Ry]&$E_{\nu,35}$[Ry]\\
\hline
0 &-1.19498    &            &-1.18715    &-1.16765    &-1.13954    &-1.10638    &-1.07137    &-1.03707    &-1.00541\\
  &-1.194278126&-1.194791662&-1.186463458&-1.167016413&-1.139029777&-1.106023726&-1.071105378&-1.036724792&-1.004827669\\
1 &-1.17484    &            &-1.16742    &-1.14898    &-1.12248    &-1.09134    &-1.05859    &-1.02662    &\\
  &-1.174311358&-1.174816204&-1.166909592&-1.148504500&-1.122055804&-1.090936653&-1.058146451&-1.026104261&\\
2 &-1.15593    &            &-1.14892    &-1.13152    &-1.10659    &-1.07738    &-1.04676    &-1.01707    &\\
  &-1.155503809&-1.156001122&-1.148500959&-1.131104266&-1.106150004&-1.076878002&-1.046197512&-1.016526876&\\
5 &-1.10619    &            &-1.10034    &-1.08586    &-1.06526    &-1.04140    &-1.01701    &            &\\
  &-1.105681500&-1.106162942&-1.099812309&-1.085293962&-1.064648793&-1.040804555&-1.016540673&&\\
10&-1.04386    &            &-1.03977    &-1.02985    &-1.01634    &-1.00223    &&&\\
  &-1.043396739&-1.043873998&-1.039336482&-1.029478689&-1.016054657&-1.001994126&&&\\
15&-1.00744    &            &-1.00544    &-1.00117    &&&&&\\
  &-1.007190171&-1.007694280&-1.005167397&-1.000829395&&&&&\\
18&-1.00017    &            &            &&&&&&\\
  &-0.999674864&-1.000213178&            &&&&&&\\
19& \,\,------                &            &            &&&&&&\\
  &-0.999462461&-1.000006426&            &&&&&&\\
\hline\hline
\end{tabular}
\end{center}}
\end{table}

\begin{figure}[!thb]
\includegraphics[scale=1.1]{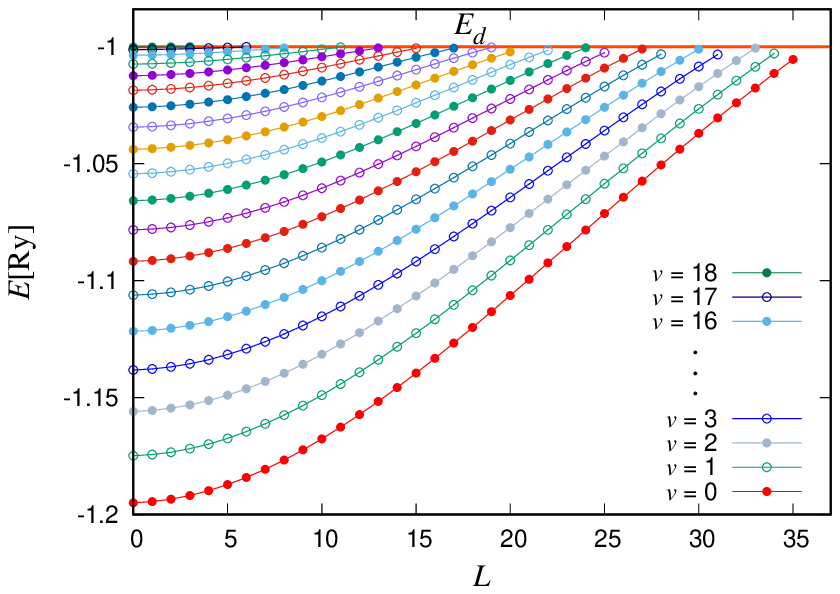}
\caption{Rovibrational bound states supported in the ground state $1s\si_g$ of the molecular ion
H$_2^+$ in the Born-Oppenheimer approximation. Each point represents the state with angular momentum
$L$ and vibrational quantum number $\nu$.  The dissociation energy $E_d = -1.0$~Ry. }
\label{rovibS}
\end{figure}

\begin{figure}[!thb]
\includegraphics[scale=1.4]{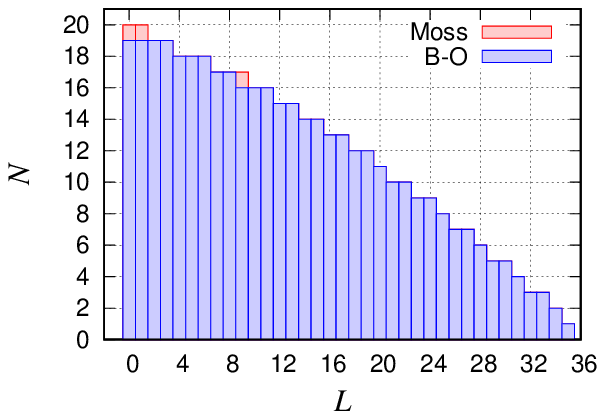}
\caption{Number of rovibrational bound states in the ground state $1s\si_g$ of the molecular ion
H$_2^+$ as a function of the angular momentum $L$. The three extra bound states found beyond B-O approximation and reported in~\cite{M:1993} are indicated in red.}
\label{rovibSH}
\end{figure}

Figure~\ref{rovibS} displays the 420 rovibrational bound states supported by the ground state potential calculated in the Lagrange mesh method. Note that beyond the Born-Oppenheimer approximation
in~\cite{M:1993} it is reported 423 bound states.
The total number of vibrational bound states for each value of the angular momentum $L$ is
presented by the histogram on Figure~\ref{rovibSH}.

\subsubsection{The first excited state $2p\si_u$}

In the same way as for the ground state $1s\si_g$ one can analyze the first excited state $2p\si_u$
using for the analytic potential ~\re{pt2p} in framework of the equation \re{vrs}.  It is well known from the highly-accurate calculations that the $2p\si_u$  potential curve at the large internuclear distance $R=12.545\,25$~a.u. displays a shallow minimum, $E_t = -1.000\,122$~Ry\,,~see e.g.\cite{Turbiner:2011}.
Taking the first derivative of our approximate potential curve~\re{pt2p} equal to zero it is found a minimum at $R=12.7034$~a.u. with $E_t=-1.000\,114$~Ry.  Although the position of the minima is determined with relative accuracy $\sim 10^{-2}$, the depth of the well is found with relative accuracy $\sim 10^{-5}\,$.
The vibrational and rotational states supported by this shallow minima can be calculated by solving~\re{vrs}.  Table~\ref{t2psuv} presents the results obtained for the three rotational states $L=0,1,2$ with vibrational quantum number $\nu=0$, calculated with the proton reduced mass $\mu=918.048$. No states with $\nu>0$ are seen even though in~\cite{P:1969}
is presented a state with $\nu = 1$, this is beyond the accuracy of the Bohr-Oppenheimer approximation.
Results by Peek~\cite{P:1969} and by Moss~\cite{M:1993} are presented for comparison. Evidently even
taking the most accurate value of the proton to electron mass ratio $\mu= 918.076336945$ does not change
our agreement with \cite{P:1969}. The presented spectra of these states agree with results in \cite{P:1969} within six figures and correspond to weak-bound states with extremely-small binding energy of $10^{-5} - 10^{-6}$\,a.u. These results for energies are definitely beyond the domain of applicability of the Born-Oppenheimer approximation: finite-mass corrections change the fifth figure, see Table~\ref{t2psuv}. Thus, the question about the existence of three rotational bound states associated with $2p\sigma_u$ potential curve remains open in the Born-Oppenheimer approximation. However,
taking into account the finite-mass effects \cite{M:1993} leads to the conclusion that three rotational bound states $L=0,1,2$ with vibrational quantum number $\nu=0$ do exist. It is quite surprising that so much shallow well keeps vibrational bound states.

\begin{center}
\begin{table}
\caption{Vibrational ($\nu$) and rotational energies in Ry ($L$) for the first electronic excited state $2p\si_u$. Comparison with Peek~\cite{P:1969} and those for the full geometry three body system given by Moss~\cite{M:1993} presented (rounded)}
\label{t2psuv}
\begin{tabular}{lllll}
\hline\hline
$\nu$ & $L$ &\ $E_{\nu L}$[Ry] \ &\qquad \cite{P:1969} &\qquad \cite{M:1993}\\
\hline
0 & 0 &-1.00003 & -1.000031268 & -0.999487004 \\
0 & 1 &-1.00002 & -1.000022797 & -0.999478536 \\
0 & 2 &-1.00001 & -1.000007307 & -0.999463045 \\
$E_{diss}$&&-1.0& -1.0                 &-0.999455679\\
\hline\hline
\end{tabular}
\end{table}
\end{center}

\subsection{The excited states $2p \pi_u$ and $3d \pi_g$}

Let us consider now the excited states $2p\pi_u$ and $3d\pi_g$. Interestingly, at large internuclear distances these states are almost degenerate.

The expansion of the energy gap $\De E\ =\ E_{3d\pi_g}\ -\ E_{2p\pi_u}$ between these states
for small $R$ was found long ago~\cite{B:1958}
\begin{equation}
\label{DE2-0}
  \De E =  \frac{5}{9}-\frac{197}{2835}\,R^2 + \frac{29190487}{281302875}\,R^4+ O\,(R^5)\ ,
\end{equation}
while at large $R$ the corresponding expansion~\cite{Cizek:1986} is
\begin{equation}
\label{DE2-infty}
  \De E\ =\ \frac{1}{2}\, R^2\, e^{-R/2-2}\left(1 + \frac{6}{R} -\frac{40}{R^2} + \cdots\right)\ + O(e^{-R}).
\end{equation}

Taking numerical data for potential curves of the $2p\pi_u$ and $3d\pi_g$ states (see Table 3 in~\cite{OT:2016})
we calculate the difference $\De E$. This difference is a smooth,
slow-changing curve with $R$, exponentially-vanishing at $R \rar \infty$, see below
Fig.~\ref{e0anddE}. Let us make interpolation between small and large distances
using the (modified) Pad\'e-type approximation $e^{-R/2-2}\ \mbox{Pade}[N+2/N](R)$ taking for a future convenience $N=7$,
\begin{equation}
\label{f22}
      \De E \ =\ e^{-R/2-2}\ \frac{5\,e^2+a_1 R+a_2 R^2+a_3 R^3+a_4 R^4+a_5 R^5+a_6 R^6
      +R^7} {9+\al_1 R+\al_2 R^2+\al_3 R^3+\al_4 R^4+2 R^5}\ ,
\end{equation}
with four constraints
\begin{eqnarray}
   \alpha_1&= &(18\,a_1-45\,e^2)/(10\,e^2)\ ,\nonumber\\
   \alpha_2&= &(-36\,a_1+72\,a_2 +3151\,e^2/35)/(40\,e^2)\ ,\nonumber\\
   \alpha_3&=&(152+2\,a_5-12\,a_6)\ ,\\
   \alpha_4&=&(-12+2\,a_6)\ ,\nonumber
\end{eqnarray}
imposed. These constraints guarantee that the appropriate expansions of (\ref{f22}) reproduce correctly the $R^0$, $R^1$ and $R^2$ terms in (\ref{DE2-0}) and the three terms in (\ref{DE2-infty}). After making a fit of the numerical data with (\ref{f22}), the six free parameters are fixed
\begin{equation}
\label{parmsfits2}
\begin{array}{ll}
a_1 = 17464.\ ,\  &  a_4 = 422.66\ , \\
a_2 = 6995.2\ ,\  &  a_5 =23.714\ , \\
a_3 = 11.559\ ,\  &  a_6=-5.7896\ ,
\end{array}
\end{equation}
where all five figures shown are significant.
This fit gives, in general, 4-5\,figures at the whole range $R \in [0,40]$~a.u., see Table \ref{tede2}, and
up to 8\,d.d. for large $R \in [20,40]$\,a.u. \ (see for illustration Table~\ref{tede2} and Fig.~\ref{e0anddE}).

\begin{center}
\begin{table}
\caption{$E_0$ and $\Delta E$ for the two excited states $E_{2p\pi_u}$ and  $E_{3d\pi_g}$ as a function of the internuclear distance $R$ compared to the results of the fits~\re{e0R2p1}~and~\re{f22}, respectively.}
\label{tede2}
\begin{tabular}{r| ll | ll}
\hline\hline
  &\multicolumn{2}{c}{$E_0=\frac{1}{2}(E_{3d\pi_g}+E_{2p\pi_u})+\frac{1}{4}$}\vline& \multicolumn{2}{c}{$\Delta E=E_{3d\pi_g}-E_{2p\pi_u}$}\\ \cline{2-5}
  R& \qquad data &Fit~\re{e0R2p1}& \qquad data &Fit~\re{f22}\\
\hline
 0.1&19.5280955405& 19.528 128&  0.5548636236&    0.554815 \\
 0.5& 3.5350863839&  3.535 089&  0.5395466409&    0.539364 \\
 1.0& 1.5523354867&  1.552 324&  0.5011027983&    0.501303 \\
 2.0& 0.5945285535&  0.594 482&  0.4041443865&    0.404110 \\
 3.0& 0.3005373017&  0.300 585&  0.3135166791&    0.313372 \\
 4.0& 0.1682219136&  0.168 226&  0.2397424036&    0.239864 \\
 5.0& 0.0985013012&  0.098 458&  0.1825418518&    0.182640 \\
 6.0& 0.0588028819&  0.058 783&  0.1389053464&    0.138851 \\
 7.0& 0.0352842264&  0.035 295&  0.1056662185&    0.105586 \\
 8.0& 0.0210978504&  0.021 103&  0.0802382260&    0.080245 \\
 9.0& 0.0125143689&  0.012 495&  0.0606848646&    0.060762 \\
10.0& 0.0073668353&  0.007 341&  0.0455988312&    0.045662 \\
20.0& 0.0005399871&  0.000 502&  0.0014141722&    0.001382 \\
30.0& 0.0002547375&  0.000 243&  0.0000212860&    0.000021 \\
40.0& 0.0001272650&  0.000 125&  2.498772$\times 10^{-7}$&2.487$\times 10^{-7}$\\
\hline\hline
\end{tabular}
\end{table}
\end{center}

\subsubsection{The dissociation energies}

The behavior of the potential curve $E_{2p\pi_u}$ at the two asymptotic limits of small and large distances is well known.  For $R\rightarrow 0$ the dissociation energy is given
by~\cite{B:1958}
\begin{equation}
\label{R0-2}
    {\tilde E}_{2p\pi_u}^{(0)}\ =\ \frac{2}{R}\ -\frac{3}{4}\ +\ \frac{1}{15}\,R^2\ -\ \frac{2447}{23625}\,R^4\ +\ \cdots \ ,
\end{equation}
while for  $R\rightarrow \infty$ the expansion reads~\cite{OV:1964,DP:1968,Cizek:1986}
\begin{eqnarray}
\label{vinfg1}
  {\tilde E}_{2p\pi_u}^{(\infty)} & = & \frac{12}{\, R^3}\ -\ \frac{156}{R^4}\ +\ \frac{4800}{R^6}\ +
  \cdots\\\non
  && - \frac{1}{4}R^2\,e^{-R/2-2}\left[1+\frac{6}{R}-\frac{40}{R^2}-\frac{940}{3\, R^3}
  -\frac{363}{R^4}\ +\ \cdots\right]\ +\ O(e^{-R})\ .
\end{eqnarray}
Considering now the negative parity state $3d\pi_g$, its expansion for small and large $R$-distance
are also well known,
\begin{equation}
\label{R0-4}
    {\tilde E}_{3d\pi_g}^{(0)}\ =\ \frac{2}{R}\ -\frac{7}{36}- \frac{8}{2835}R^2\ +\frac{54058}{281302875}R^4 +\ \ldots\ ,
\end{equation}
at $R \rar 0$, where the  term  $O(R)$ is absent.
The behavior for $R \rar \infty$ is the same as one given by Eq.~\re{vinfg1} with sign changed from minus to plus in front of the exponentially-small term $\sim e^{-R/2}$.

Let us consider the half-sum of potential curves for
$2p\pi_u$ and $3d\pi_g$ states,
\begin{equation}
\label{E1-sum}
     E_0\ \equiv \ \frac{{\tilde E}_{2p\pi_u}+{\tilde E}_{3d\pi_g}}{2}\ .
\end{equation}
Its corresponding expansions are
\begin{equation}
\label{R1-sum}
 E_0\ =\ \frac{2}{R}\ -\ \frac{17}{36}\ +\ \frac{181}{5670}R^2 +O(R^4)\ ,
\end{equation}
at $R \rar 0$, where the linear in $R$ term is absent, and
\begin{equation}
\label{Rinfty1-sum}
 E_0\ =\ \frac{12}{ R^3}\ -\ \frac{156}{R^4}\ +\ \frac{4800}{R^6}\ +\ \ldots\
 +\ O(e^{-R})\ ,
\end{equation}
at $R \rar \infty$, where again the term $R^{-5}$ is absent. The first expansion (\ref{R1-sum})
has a finite radius of convergence, see e.g. \cite{K:1983}, while the second one (\ref{Rinfty1-sum}) (the half-sum of two multipole expansions) has zero radius of convergence.

Assuming that two-instanton contribution, $\sim e^{-R}$ a large $R$ can be neglected,
we construct an analytic approximation for $E_0$ which mimics the two asymptotic limits
(\ref{R1-sum}), (\ref{Rinfty1-sum}). The most convenient way to do it is to use Pad\'e type
approximations $E_0(R)=\frac{1}{R}\ \mbox{Pade}[N/N+2](R)$ with a certain $N$.
Concrete fit was made for $N=7$, where the Pad\'e approximation is of the form
$E_0(R)=\frac{1}{R}\ \mbox{Pade}[7/9](R)$,
\begin{equation}
\label{e0R2p1}
   E_0 =\frac{2+ R+a_2R^2 +a_3R^3 +a_4R^4 +a_5R^5+a_6R^6+12R^7}{R (1 +\al_1R
  +\al_2R^2+b_3R^3+b_4R^4+b_5R^5+b_6R^6+\al_3R^7+\al_4R^8+R^9)}\ ,
\end{equation}
with four constraints imposed,
\begin{equation}
\begin{array}{l}
   \mbox{$\al_1= 53/72$}\ ,\\
   \mbox{$\al_2= (901+2592a_2)/5184$}\ ,\\
   \mbox{$\al_3= (a_5+13a_6+2028)/12$}\ ,\\
   \mbox{$\al_4= (a_6+156)/12$}\ ,
\end{array}
\end{equation}
in order to guarantee that the three coefficients in front of $R^{-1}$, $R^{0}$ and $R^{1}$ of expansion
at $R\rightarrow 0$, see \re{R1-sum} and the three coefficients in front of $R^{-3}$, $R^{-4}$ and
$R^{-5}$ at $R\rightarrow \infty$, in the $1/R$-expansion (\ref{Rinfty1-sum}) are exact.
Thus, (\ref{e0R2p1}) depends on nine free parameters. After making the fit with~(\ref{e0R2p1}),
we arrive to concrete values of these nine free parameters:
\begin{equation}
\label{parmsfits3}
\begin{array} {lrlr}
a_2 =&  677043.\ ,  & b_3 =&  107330.\ ,\\
a_3 =&  54874.3\ ,  & b_4 =&  21319.7\ ,\\
a_4 =& -8925.37\ ,  & b_5 =& -839.229\ ,\\
a_5 =&  2332.69\ ,  & b_6 =& 703.849\ , \\
a_6 =& -296.134\ . \\
\end{array}
\end{equation}
It provides not less than \,4-6\,figures in $E_0$ for whole studied domain $R \in [1,40]$~a.u.
(see Table~\ref{tede2}  and Fig.~\ref{e0anddE}).

The potential curve for the $2p\pi_u$ state can be constructed from \re{e0R2p1} and \re{f22} by taking
\begin{equation}
\label{f2ppu}
   E_{2p\pi_u}\ =\ E_0 - \frac{1}{2}\De E\ .
\end{equation}
This expression reproduces 4-5-6 figures in energy for the whole domain $R \in [0, 40]$~a.u.,
when comparing with the exact values (see Table 3 in \cite{OT:2016}). For illustration see
Fig.~\ref{pch2pRy}. The fit predicts the minimum of the
potential curve as  $E_{min} = -0.269\,022$\,Ry (cf. $E_{min}^{exact}=-0.269\,027\,6$\,Ry (rounded)) while its
location is  $R_{eq}=7.957$\,a.u. (c.f. $R_{eq}^{exact}= 7.931$\,a.u.). Due to the fact that dominant long range interaction is repulsion $\sim 12/R^3$ (see~\re{vinfg1}) the potential curve should display a maximum. Again, it is noticeable that the fit predicts the maximum of the potential curve as  $E_{max} = -0.249\,734$\,Ry (c.f. $E_{max}^{exact}=-0.249\,713$\,Ry (rounded)) which is localized at  $R_{max}=26.138$\,a.u.
(c.f. $R_{max}^{exact}= 25.809$\,a.u. (rounded)).


The asymptotic expansions of Eq.~(\ref{f2ppu})
are given by
\begin{eqnarray}
      &{\tilde E}_0^{}&=\frac{2}{R}\ -\frac{3}{4} +71.8\,R^2\ +\cdots \ ,\\
      &{\tilde E}_{\infty}&=\frac{12}{R^3}-\frac{156}{R^4}-\frac{10729.2}{R^6}+\cdots
      -\frac{1}{4}R^2e^{-R/2-2}\left[ 1 + \frac{6}{R}-\frac{40}{R^2}-\frac{645.2}{R^3}+\cdots\right] \ ,
\end{eqnarray}
which are in complete agreement with the first three terms at
$R\rightarrow 0$ and $R\rightarrow \infty$ (in the $1/R$ expansion of the pre-factor to
$e^{-R/2}$, cf.~\re{R0-2} and \re{vinfg1}).

Similarly, the potential curve for the excited state $3d\pi_g$ is
restored from \re{e0R2p1} and \re{f22} by taking
\begin{equation}
\label{f3dpg}
   E_{3d\pi_g}\ =\ E_0 + \frac{1}{2}\De E\ .
\end{equation}
This expression also reproduces 4-6 figures when comparing with the exact energy, see Table 3 in \cite{OT:2016} and for illustration Fig.~\ref{pch2pRy}. The asymptotic expansions of Eq.~(\ref{f3dpg})
are given by
\begin{eqnarray}
     &{\tilde E}_0^{}&=\frac{2}{R} -\frac{7}{36}+ 71.7R^2+\cdots \ ,\\
     &{\tilde E}_{\infty}&=\frac{12}{R^3}-\frac{156}{R^4}-\frac{10729.2}{R^6}+\cdots
      +\frac{1}{4}R^2e^{-R/2-2}\left[ 1 + \frac{6}{R}-\frac{40}{R^2}-\frac{645.2}{R^3}+\cdots\right] \ ,
\end{eqnarray}
which are in complete agreement with the first three terms at
$R\rightarrow 0$ (cf.~\re{R0-4}), and in both three terms in $1/R$ expansion
and three terms in $1/R$ expansion of the pre-factor to $e^{-R/2}$ for
$R\rightarrow \infty$ (cf.~\re{vinfg1}). This potential curve displays no minimum, thus,
the state $3d\pi_g$ is pure repulsive.

\subsubsection{Vibrational states associated with the potential curve $2p\pi_u$}

\begin{figure}[!thb]
\includegraphics[scale=1.4]{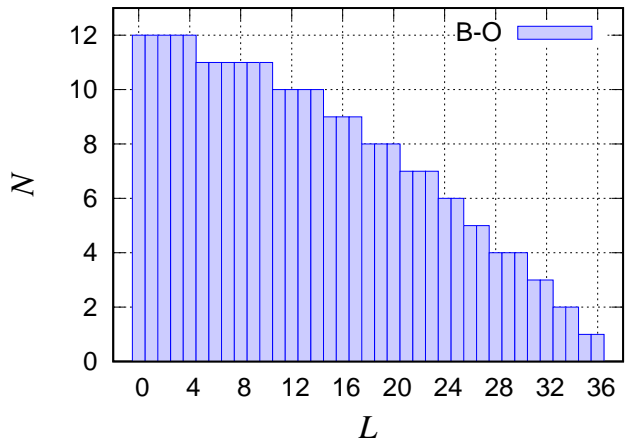}
\caption{H$_2^+$:\ Number of vibrational states supported by the excited state $2p\pi_u$ as a function of
the angular momentum $L$, here $L_{max}=36$. In total, there are 284 rovibrational states.}
\label{rovib2ppu}
\end{figure}

Taking the function~\re{f2ppu} as the potential $V(R)$ for the  $2p\pi_u$-state, we
calculate the rovibrational bound states by solving~\re{vrs} at different $L$.  In order to solve this one-dimensional differential equation the Lagrange-mesh method is used~\cite{Baye:2015} as for $1s\si_g$. Table~\ref{t2ppuv} presents some of the obtained rovibrational states for $L=0, 4$, where the reduced mass of the two protons was taken equal to $\mu=918.048$.
This potential well keeps 12 vibrational states at $L = 0$ and 284 rovibrational states in
total, see Figure \ref{rovib2ppu}. Beckel el al. \cite{BSP:1973} reported vibrational states for
$L = 1, 4, 8, 12, 16$ where the first correction to the Born-Oppenheimer approximation was included.
At $L=4$ our results differ from \cite{BSP:1973} systematically at 5th figure which likely the order of
the contribution of the first correction to the Born-Oppenheimer approximation.
Maximum angular momentum $L$, which still keeps vibrational bound state, is equal to $L_{max}=36$.

We are not aware about any calculations beyond of the Bohr-Oppenheimer approximation for the state $2p\pi_u$. It seems likely that some rovibrational states can continue to exist.

\begin{center}
\begin{table}
\caption{H$_2^+$:\ Rovibrational energies $E_{\nu L}$ of the excited state $2p\pi_u$ for $L=0, 4$.
On second line for $L=4$ case the results from~\cite{BSP:1973}. }
\label{t2ppuv}
\begin{tabular}{rll}
\hline\hline
$\nu$&$E_{\nu0}$[Ry]&$E_{\nu4}$[Ry]\\
\hline
0  & -0.267\,83&-0.267\,50\\
   &           &-0.267\,507\,4\\
1  & -0.265\,54&-0.265\,22\\
   &           &-0.265\,214\,9\\
2  & -0.263\,35&-0.263\,04\\
   &           &-0.263\,043\,8\\
5  & -0.257\,52&-0.257\,25\\
   &           &-0.257\,299\,3\\
10 & -0.250\,87&-0.250\,71\\
   &           &-0.250\,759\,0\\
11 & -0.250\,16 &-0.250\,04\\
   &           &-0.250\,052\,4\\
\hline\hline
\end{tabular}
\end{table}
\end{center}

\section{The ground state X$^2\Si^+$ of the H\lowercase{e}H molecule}

The analysis implemented above for the homonuclear molecule H$_2^+$ can also be extended to the case
of heteronuclear molecules. As an example, let us consider the simplest neutral diatomic molecule made up
of a Helium atom plus a Hydrogen atom, the (HeH) molecule. The potential curve for the ground state as a function
of the internuclear distance $E(R)$ is repulsive with a shallow van der Waals minimum at $R=6.66$~a.u.~\cite{MF:1994}, \cite{MWB:2002}. The united atom limit $R\rightarrow 0$ evidently corresponds to the Li atom while at large
internuclear distances the HeH molecule dissociates like \mbox{HeH$ \rar $ He $+$H}.
The dissociation energy is given by
\begin{equation}
\tilde{E}=E(R)- (E_{\text He}+E_{\text H}),
\end{equation}
where $E_{\text He}=-8.71017$~a.u. and $E_{\text H}=-0.99946$~a.u.~\cite{WTT:2015}.

The dissociation energy $\tilde{E}$ for small internuclear distances $R\rightarrow 0$ can be expanded
as~\cite{B:1958}
\begin{equation}
\label{heh1}
\tilde{E}\ =\ \frac{4}{R}-\varepsilon_0 + O(R^2) + \cdots\,,
\end{equation}
where the first them is the repulsive interaction $2Z_1Z_2/R$ while the term $\varepsilon_0$
corresponds to the absolute value of the united atom energy minus sum of the energies
of H and He,
$\varepsilon_0=|E_{\text{Li}}-(E_{\text He}+E_{\text H})|=8.14972$~Ry. As was pointed out long ago by
Buckingham~\cite{Bh:1958} the linear term $\sim R$ in (\ref{heh1}) should be absent. At large internuclear distances $R\rightarrow \infty$ the dissociation energy is expanded
as~\cite{YBD:1996}
\begin{equation}
\label{heh2}
\tilde{E} = -\frac{C_6}{R^6}-\frac{C_8}{R^8}-\frac{C_{10}}{R^{10}} +\cdots\,,
\end{equation}
where the (rounded) coefficients $C_i$ are $C_6=5.64268$, $C_8=83.6728$ and $C_{10}=1 743.080$, all in a.u.
For our purposes it has to be paid attention that the next-after-leading term $\sim 1/R^7$ is absent.

Now, in order to construct an analytic approximation of the potential curve which mimics the two
asymptotic limits~(\ref{heh1}) and (\ref{heh2}) a Pad\'e type (meromorphic) approximation
of the form $E(R)=\frac{1}{R}\text{Pade}[N/N+5](R)$ is used. Taking $N=4$, which seems appropriate
for our purposes, the explicit expression for $E(R)=\frac{1}{R}\text{Pade}[4/9](R)$ is
\begin{equation}
\label{heh3}
  \hspace{0.5cm} E\ =\
\frac{(4+a_1R+a_2R^2+a_3R^3-C_6R^4)}
{R(1+\al_1 R+\al_2R^2+b_3 R^3+b_4 R^4+b_5 R^5+b_6 R^6-\al_3 R^7-\al_4 R^8+R^9)}\ ,
\end{equation}
with four constraints
\begin{eqnarray}
\al_1 &=& (a_1+\varepsilon_0)/4\,,\\
\al_2 &=& (-a_1\varepsilon_0 + \varepsilon_0^2 + 4a_2)/16\,,\nonumber\\
\al_3 &=& (a_2 + C_8)/C_6\,,\nonumber\\
\al_4 &=& a_3/C_6\,.\nonumber
\end{eqnarray}
which guarantees that the coefficients in front of $R^{-1}$, $R^{0}$ and $R^{1}$ at
small internuclear distances
in~(\ref{heh1})  and the coefficients in front of $R^{-6}$, $R^{-7}$ and $R^{-8}$ for
large internuclear distances in~(\ref{heh2}) are reproduced exactly.
After making a fit with the analytic expression~(\ref{heh3}), using the
numerical energy values reported by Murrell~{\it et al.}~\cite{MWB:2002},
the values of the remaining seven free parameters are
\begin{equation}
\begin{array}{ll}
a_1 =  2072.1784, & b_3 =   2075.4703,\\
a_2 = -604.31543, & b_4 =  -2101.4132,\\
a_3 =  76.347450, & b_5 =  1152.5705,\\
                  & b_6 =  -380.63678.
\end{array}
\end{equation}

\begin{figure}
\begin{center}
\includegraphics[scale=1.2]{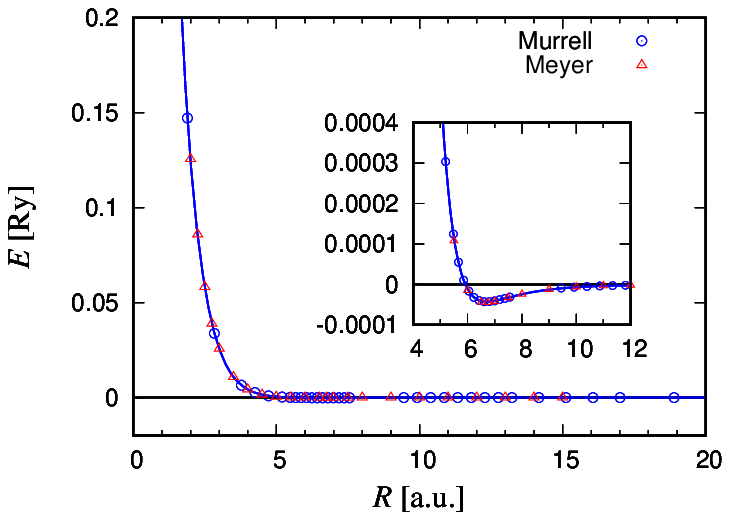}
\end{center}
\caption{HeH:\ Dissociation energy of the ground state of HeH.
 The solid curve is fit~\re{heh3}, it compared with theoretical calculations by
 Murrell~{\it et al.}~\cite{MWB:2002} and Meyer~{\it et al.}~\cite{MF:1994}.
 The van der Walls minimum (amplification) is located at $R \approx 6.7$~a.u.}
\label{f1heh}
\end{figure}

Expression~\re{heh3} provides 6-9 figures in the total energy as can be seen in Table~\ref{t1heh}.
Taking the derivative of~\re{heh3} the minimum of the potential curve is predicted as
$E_{min}^{(fit)}=-6.806\,28$~Ry  (c.f. $E_{min}=6.806\,29$~Ry from~\cite{MF:1994})
at $R_{min}^{(fit)}=6.707$~a.u. (c.f. $R_{min}=6.660$~a.u. from~\cite{MF:1994}).
Figure~\ref{f1heh} displays the analytic potential~\re{heh3} as well as the theoretical energy
values used to make the fit~\cite{MWB:2002} (blue circles). Comparison with results by
Meyer~{\it et al.}~\cite{MF:1994} (red triangles) are also depicted.
The small plot in Figure~\ref{f1heh} amplifies the details of the curve
around the shallow minimum.

\begin{table}
\caption{HeH:\ Total energy $E_0$ for the ground state {\it vs.} $R$ in a.u.
Second column corresponds to the results presented in \cite{MWB:2002} and
third column are the results of the fit~\re{heh3}.}
\label{t1heh}
\begin{center}
\begin{tabular}{r| rr}
\hline\hline
&\multicolumn{2}{c}{$E_0\times 10^{-6}$[Ry]}\\
\hline
$R$\ &\ data &\ Fit\\
\hline
 1.890& 147183.334 \ & 147183.306 \\
 2.835&  33875.651 \ &  33875.635 \\
 4.252&   2633.407 \ &   2633.024\\
 6.047&    -16.548 610&    -16.299\\
 6.236&    -31.985 473&    -31.646\\
 6.425&    -39.813 257&    -39.475\\
 6.614&    -42.765 763&    -42.492\\
 6.803&    -42.674 636&    -42.528\\
 6.992&    -40.806 539&    -40.800\\
 7.181&    -37.954 273&    -38.107\\
10.393&     -5.075 757&     -5.474\\
12.283&     -1.795 196&     -1.934\\
14.173&     -0.738 126&     -0.790\\
16.063&     -0.346 281&     -0.362\\
18.897&     -0.127 577&     -0.133\\
22.677&     -0.041 918&     -0.0434\\
26.456&     -0.016 403&     -0.0170\\
30.236&     -0.007 290&     -0.0076\\
37.795&     -0.001 823&     -0.00196\\
\hline\hline
\end{tabular}
\end{center}
\end{table}


The asymptotic expansions of Eq.~\re{heh3} are given by
\begin{eqnarray}
E_{0} &=&\frac{4}{R} - 8.14972- 821.0904~R^2+\cdots\,\\
E_{\infty} &=& -\frac{5.64268}{R^6}-\frac{83.6728}{R^8}-\frac{1207.756}{R^9}+\cdots \ ,
\end{eqnarray}
respectively,
which reproduce the first three term of $R^{-1}$, $R^0$ and $R$ for $R\rightarrow 0$ (see \re{heh1})
and the first three terms $R^{-6}$, $R^{-7}$ and $R^{-8}$ for $R\rightarrow \infty$  (see \re{heh2}).

The traditional molecular system (HeH) has been studied numerically for a long time,
see Murrell~{\it et al.}~\cite{MWB:2002} and references therein. There are always a certain
controversies between different calculations. E.g. for total energy at $R=6.0$\,a.u.
data by Meyer~{\it et al.}~\cite{MF:1994} differ from Murrell~{\it et al.}~\cite{MWB:2002} with the
relative difference $\sim 30\%$. Although for other values of $R$ the relative difference for these two calculations does exceed $\lesssim 7\,\times \, 10^{-2}$. Making fit to find parameters in (\ref{heh3})
we used data by Murrell~{\it et al.}~\cite{MWB:2002}, while excluding the point $R=6.0$\,a.u.

Recently, a certain fit of the potential curve was proposed by Warnicke~{\it et al.}~\cite{ WTT:2015}.
When comparing our fit~\re{heh3} with the fit $V_{TT2}$ \cite{ WTT:2015}, the relative difference
goes from $\sim 1\%$ for large $R$ up to $\sim 30\%$ for small distances, $R \sim 1.$\,a.u. while being reasonably small at intermediate $R$.
The main deficiency of the $V_{TT2}$ potential is the wrong description
of the asymptotic behavior for both $R\rightarrow 0$ and $R\rightarrow \infty$, as one can see that
by expanding $V_{TT2}$ at small and large distances.

The analytic expression for the potential curve of the ground state X$^2\Si^+$ together with~\re{vrs}
allows us to calculate possible rovibrational states supported by the shallow minimum.
Using the Lagrange-mesh method to solve~\re{vrs} we claim that the ground state does not keep any
rovibrational state(!) confirming what was already mentioned in numerical studies, see \cite{MWB:2002}.


\section{The hydrogen molecule H$_2$}

The next system to consider is the homonuclear hydrogen molecule H$_2$. It is well known that the potential curve of the ground state $^1\Si_g^+$ has well-pronounced, profound minimum at $R=1.4011$~a.u.~\cite{P:2010}, while the first excited state $^3\Si_u^+$  exhibits a shallow minimum at $R=7.85$~a.u.~\cite{KW:1965} being repulsive. In the limit of small internuclear distances $R\rightarrow 0$ this molecule is converted to the helium atom (the so-called united atom limit)\footnote{If interproton repulsion term $1/R$ should be removed}. At large internuclear distances $R\rightarrow \infty$ the molecule dissociates into two hydrogen atoms, H$_2\rightarrow$ H+H (it is the main dissociation channel).

\subsection{Energy gap (exchange energy)}

It can be easily seen that the energy gap between the states $^3\Si_u^+$ and $^1\Si_g^+$,
\begin{equation}
\label{deh2exp}
    \De E\ =\ E_{^3\Si_u^+}\ -\ E_{^1\Si_g^+}\ ,
\end{equation}
at small internuclear distances behaves as
\begin{equation}
\label{h2d0}
     \De E^{(0)}\ =\ \de_0 + 0 \cdot R + O(R^2)\ ,
\end{equation}
where $$\de_0 = E_{2^3S}^{He} - E_{1^1S}^{He}\ ,$$ is the energy difference between the excited $2^3S$ and the ground $1^1S$ states of the (united) helium atom~\cite{HB:2001,GD:1996}
\begin{eqnarray*}
 E_{1^1S}^{He} & = & -5.807\,448\,754\,068~\mathrm{Ry}\ , \\
 E_{2^3S}^{He} & = & -4.350\,458\,756\,473~\mathrm{Ry}\ .
\end{eqnarray*}
At large internuclear distances there exists a controversy in literature about the behavior of $\De E$  \footnote{in this domain $\De E$ is called {\it exchange energy}}
~\cite{HF:1964,KY:1967,TTY:1993,BDC:2010}. We made a choice of the behavior derived in \cite{HF:1964} and confirmed at \cite{LL}, Ch.XI, p.315, 
\begin{equation}
\label{h2di}
\Delta E^{(\infty)} =  R^{b}\, e^{-a\,R}\left(\ep_0\ +\ \frac{\ep_1}{R}\ +\ \frac{\ep_2}{R^2} + \cdots\right),
\end{equation}
where $a=2, b=5/2$ and $\ep_0=3.28$\,Ry, and, in general, \{$\ep_i$\} are parameters to be specified \footnote{The parameters $a,b, \ep_0$ are evidently related with instanton action and determinant in instanton calculus. We are not aware about the reliable calculations of these quantities.}.
Numerically, $\De E$ is calculated by using data of the potential curves found in ~\cite{P:2010,PK:2011}.

\begin{figure}[!thb]
\includegraphics[scale=1.4]{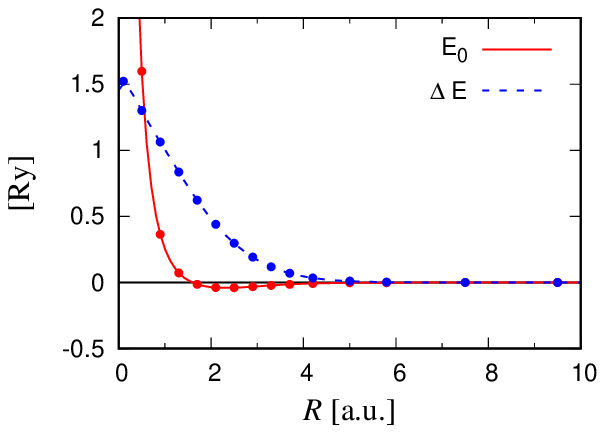}
\caption{$E_0$ and $\De E$ for $^1\Si_g^+$ and  $^3\Si_u^+$ states as defined by~\re{e0h2exp}
  and~\re{deh2exp},  respectively. Calculated energies are marked by dots, the solid curves
  are fits~\re{h2E0fit} (solid line) and \re{h2dEfit} (dashed line).}
\label{e0dEh2}
\end{figure}

In order to interpolate between the two asymptotic regimes \re{h2d0} and \re{h2di}, we introduced a new variable 
\[
     r\ =\ \sqrt R\ ,
\]
and construct the (modified) Pad\'e-type approximation
\[
     e^{-2 r^2}\, Pade[N + 5/N](r) \ =\ e^{-2 r^2}\,\frac{P_{N+5}(r)}{Q_N(r)}\ ,
\]
where $P_{N+5}(r), Q_N(r)$ are polynomials of degrees $(N+5), N$, respectively. To be concrete we choose
$N = 11$,
\begin{equation}
\label{h2dEfit}
 \De E \ =\ e^{-2 r^2} \frac{\de_0+(\sum_{i=1}^{8} a_i r^{2i} )}{1+\alpha_1 r^2 + (\sum_{i=2}^{4}b_i r^{2i})+ r^{11}}\ ,
\end{equation}
which provides eventually the accuracy we are looking for; here the parameter
\begin{displaymath}
   \al_1= (a_1-2\de_0)/\de_0\ ,
\end{displaymath}
is chosen in such a way to ensure the correct two terms in the expansion at $R\rightarrow 0$ (\ref{h2d0}).  $b_4=b_5=0$ guarantee the right behavior at $R\rightarrow \infty$ (\ref{h2di}). After fitting energy gap $\De E$, which is known numerically \cite{P:2010,PK:2011}, by requiring minimal $\chi^2$, the remaining 10 free parameters in (\ref{h2dEfit}) are found
\begin{equation}
\begin{array}{ll}
a_1 =  258.9433\ ,& a_6 =  82.43842\ ,\\
a_2 =  835.6547\ ,& a_7 = -9.777265\ ,\\
a_3 = -115.8885\ ,& b_8 =  3.612628\ ,\\
a_4 =  1084.584\ ,& b_2 =  91.55477\ ,\\
a_5 = -301.2744\ ,& b_3 = -21.42445\ .\\
\end{array}
\end{equation}
This fit gives, in general, 5-4-3 figures at the whole range $R \in [0,20]$ a.u., see Table~\ref{h2e0decomp} and Figure~\ref{e0dEh2}, in such a way that the number of correct figures reduces with increase of $R$. This result is especially impressive for $R \simeq 15$\,a.u. where the energy gap is already exponentially-small.

\begin{center}
\begin{table}
\caption{$E_0$ and $\De E$ (in Ry) for the two states $^1\Sigma^+_g$ and  $^3\Sigma^+_u$
 of the molecule H$_2$ as a function of the internuclear distance $R$ (in a.u.)
 compared to the results of the fits~\re{h2E0fit}~and~\re{h2dEfit}, respectively.}
\label{h2e0decomp}
\begin{tabular}{r| ll | ll}
\hline\hline
  &\multicolumn{2}{c}{$E_0=\frac{1}{2}(\tilde{E}_{^3\Sigma^+_u}+\tilde{E}_{^1\Sigma^+_g})$}\vline&
    \multicolumn{2}{c}{$\Delta E=E_{^3\Sigma^+_u}-E_{^1\Sigma^+_g}$}\\ \cline{2-5}
  R& \qquad data~\cite{P:2010,PK:2011} &Fit~\re{h2E0fit}& \qquad data~\cite{P:2010,PK:2011} &Fit~\re{h2dEfit}\\
\hline
 0.1& 17.015 417 21 & 17.003 669 2 &  1.521 967 50 & 1.501 373 2 \\
 0.2&  7.134 563 80 &  7.134 563 8 &  1.477 914 42 & 1.477 914 5 \\
 0.5&  1.596 841 73 &  1.595 420 7 &  1.300 238 49 & 1.298 101 5 \\
 1.0&  0.253 195 85 &  0.253 195 9 &  1.004 550 58 & 1.004 550 7 \\
 2.0& -0.035 209 29 & -0.035 209 3 &  0.482 113 25 & 0.482 113 5 \\
 3.0& -0.029 341 32 & -0.029 341 3 &  0.170 622 44 & 0.170 622 8 \\
 4.0& -0.009 770 32 & -0.009 770 4 &  0.046 020 37 & 0.046 020 3 \\
 5.0& -0.002 472 92 & -0.002 473 0 &  0.010 196 80 & 0.010 196 4 \\
 6.0& -0.000 649 16 & -0.000 649 1 &  0.002 044 52 & 0.002 044 7 \\
 7.0& -0.000 201 92 & -0.000 202 0 &  0.000 387 82 & 0.000 388 6 \\
 8.0& -0.000 075 83 & -0.000 075 8 &  0.000 070 77 & 0.000 071 2 \\
 9.0& -0.000 033 30 & -0.000 033 3 &  0.000 012 53 & 0.000 012 7 \\
10.0& -0.000 016 43 & -0.000 016 4 &  2.163 656e-6 & 2.202 1e-6	 \\
15.0& -1.250 780e-6 & -1.245 4 e-6 &  2.58e-10     & 2.70e-10	 \\
20.0& -2.134 80 e-7 & -2.129 8 e-7 &  0.0          & 0.0	 \\\hline\hline
\end{tabular}
\end{table}
\end{center}

\subsection{The dissociation energies}

For asymptotically large internuclear distances the molecule H$_2$ at the state $^3\Si_u^+$ or $^1\Si_g^+$ dissociates into two hydrogen atoms in its ground state of energy $E_H=-1$\,Ry. The dissociation energy is defined as
\begin{equation}
\tilde{E}\ =\ E(R)- 2 E_{\text H}\ =\ E(R) + 2\ .
\end{equation}
It is the well known that for $R\rightarrow 0$, the dissociation energy expansion for the ground state $^1\Si_g^+$ is given by~\cite{BBS:1966}
\begin{equation}
\label{h21sr0}
\tilde{E}_{^1\Si_g^+}^{(0)}= \frac{2}{R} + E_{1^1S}^{He} + 0\cdot R + O(R^2)\ ,
\end{equation}
where the linear in $R$ term is absent.
While at large internuclear distances $R\rightarrow \infty$ the dissociation energy expansion is of the form
\begin{equation}
\label{1sH2i}
\tilde{E}_{^1\Si_g^+}^{(\infty)}= -\frac{C_6}{R^6}-\frac{C_8}{R^8} + \cdots - \frac{1}{2}R^{5/2} e^{-2R}\left(\ep_0+\frac{\ep_1}{R}+\frac{\epsilon_2}{R^2} +\cdots\right),
\end{equation}
where
\begin{eqnarray}
C_6 &=&12.99805341\ ,\nonumber\\
C_8 &=&248.79816717\nonumber\ ,
\end{eqnarray}
are multipole coefficients~\cite{YBD:1996}.
On the other hand, the expansion for small internuclear distances of the excited state $^3\Si_u^+$ is given by~\cite{BBS:1966}
\begin{equation}
\label{h23sr0}
\tilde{E}_{^3\Si_u^+}^{(0)}= \frac{2}{R} + E_{2^3S}^{He} + 0\cdot R + O(R^2)\ .
\end{equation}
For large internuclear distances, the expansion is the same as for the ground state~\re{1sH2i} with different sign in front of the exponential-small term: instead of the minus sign, it should be plus~\cite{LL}.

Now, the half-sum of the potential curves for dissociation energy of these two states is
\begin{equation}
\label{e0h2exp}
   E_0\ =\ \frac{{\tilde E}_{^1\Si_g^+}+{\tilde E}_{^3\Si_u^+}}{2}\ .
\end{equation}
Its asymptotic expansions are obtained from the previous expressions (\ref{h23sr0}), (\ref{h21sr0}),
\begin{equation}
\label{e0h20}
    E_0\ =\  \frac{2}{R} + C_0+ 0 \cdot R + O(R^2) \ ,\\
\end{equation}
while for $R\rightarrow\infty$
\begin{equation}
\label{e0h2i}
    E_0\ =\ -\frac{C_6}{R^6}-\frac{C_8}{R^8} + \cdots\ ,
\end{equation}
where
\begin{equation}
    C_0\ =\ \frac{1}{2}( E_{1^1S}^{He}+ E_{2^3S}^{He}+2E_H)\ .
\end{equation}
The interpolation between the two asymptotic behaviors \re{e0h20} and \re{e0h2i} is performed using two-point Pad\'e-type approximation $Pade[N/N+5](R)/R$. In concrete consideration we choose $N = 7$. The analytic expression for $E_0$ reads
\begin{equation}
\label{h2E0fit}
   E_0\ =\ \frac{2+(\sum_{i=1}^{6}a_iR^{i}) - C_6R^7}
   {R(1+\al_1R+\al_2R^2+(\sum_{i=3}^{9}b_iR^{i}) - \al_3R^{10} - \al_4R^{11}+R^{12})}\ ,
\end{equation}
with four constraints imposed
\begin{eqnarray}
\alpha_1&=& (a_1-C_0)/2\ ,\\
\alpha_2&=& (2 a_2-a_1C_0+C_0^2)/4\ ,\nonumber\\
\alpha_3&=&(C_8+a_5)/C_6\ ,\nonumber\\
\alpha_4&=&a_6/C_6\ ,\nonumber
\end{eqnarray}
which guarantee the exact reproduction of the first two terms in both expansions (\ref{e0h20}), (\ref{e0h2i}).
$E_0$ is calculated using the numerical results for the potential curves from~\cite{P:2010,PK:2011}. After fitting with minimal $\chi^2$ the 13 free parameters for $E_0$ (\ref{h2E0fit}) are found:
\begin{equation}
\begin{array}{ll}
a_1 =    7557.03\ , & b_3 =  8555.397\ , \\
a_2 =    9880.74\ , & b_4 =  2636.858\ , \\
a_3 =  -14506.57\ , & b_5 =  6657.054\ , \\
a_4 =    5049.34\ , & b_6 =  -9627.171\ ,\\
a_5 =  -1330.003\ , & b_7 =  5257.291\ , \\
a_6 =    189.578\ , & b_8 =  -1025.438\ ,\\
                    & b_9 =  -92.54939\ .\\
\end{array}
\end{equation}
This fit gives, in general, 5-4-3 figures at the whole range $R \in [0,20]$ a.u., see Table~\ref{h2e0decomp} and Figure~\ref{e0dEh2}. Number of correct figures is reduced with increase of $R$.

\begin{figure}
\begin{center}
\includegraphics[scale=1.2]{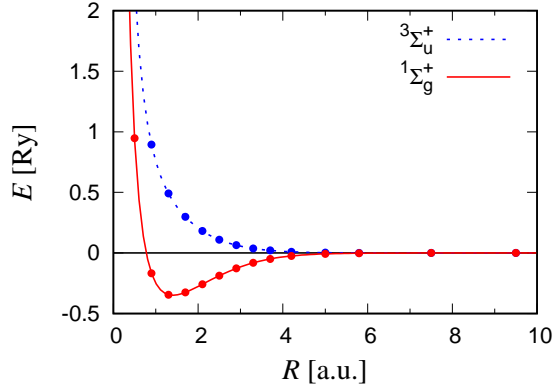}
\end{center}
\caption{Dissociation energy of the states $^1\Si_g^+$ (solid line) and $^3\Si_u^+$
  (dashed line). Points indicate data \cite{P:2010,PK:2011} and the curves represent fits.
  The minimum for the ground state $^1\Si_g^+$ is around $R \approx 1.4$~a.u.}
\label{et2pair}
\end{figure}

\subsection{Asymptotics from fits}

The potential curve for dissociation energy for the ground state $^1\Si^+_g$ can be recovered by taking
\begin{equation}
\label{h2an1s}
  {\tilde E}_{^1\Si_g^+}\ =\ E_0 - \frac{1}{2}\De E\ ,
\end{equation}
where $E_0$ and $\De E$ are given by the expressions  \re{h2E0fit} and \re{h2dEfit}, respectively.
At small internuclear distances ($R\rightarrow 0$), the dissociation energy ${\tilde E}_{^1\Si_g^+}$ takes the form
\begin{equation}
  {\tilde E}_{^1\Si_g^+}\ =\ \frac{2}{R} -5.807448754 +1417.566 R^2+\cdots,
\end{equation}
which reproduces  the coefficients in front of the terms $R^{-1}$, $R^{0}$ and $R$
in the expansion~\re{h21sr0}.
For $R\rightarrow \infty$ the expansion is
\begin{eqnarray}
\label{h2fexpri}
 {\tilde E}_{^1\Si_g^+}& = &-\frac{12.99805341}{R^6}-\frac{248.79816716}{R^8}+ \frac{217.6295}{R^9}+\cdots\\ &-&1.806\,314\,R^{5/2}e^{-2R} \left[1 -\frac{2.706413}{R}+\frac{22.8195}{R^2}+\cdots\right]\nonumber,
\end{eqnarray}
in functional agreement with the expression~\re{1sH2i}.
On the other hand, taking
\begin{equation}
\label{h2an3s}
   {\tilde E}_{^3\Si_u^+}\ =\ E_0 + \frac{1}{2}\Delta E\ ,
\end{equation}
we recover the potential curve for the excited  state $^3\Sigma^+_u$. For $R\rightarrow 0$ this expression behaves like
\begin{equation}
   {\tilde E}_{^3\Si_u^+}\ =\ \frac{2}{R}\ -\ 4.350458756\ +\ 1604.854 R^2\ +\ \cdots\ ,\\
\end{equation}
reproducing the first three coefficients of~\re{h23sr0}. At large internuclear distances the behavior is the same as that of the ground state~\re{h2fexpri} except for the opposite sign in front of the exponential-small term.

Expressions  \re{h2an1s} and \re{h2an3s} are an analytic representation of the potential curves for dissociation energy for the states  $^1\Si_g^+$ and $^3\Si^+_u$ of the hydrogen molecule H$_2$, respectively.  Taking the derivative of the expression~\re{h2an1s} and putting it to zero predicts a position of minimum for the ground state potential curve $E_t=-2.348942$\,Ry at $R=1.4012$\,a.u. while the accurate result is $E_t=-2.3489518628$\,Ry (rounded) at $R=1.4011$\,a.u., see ~\cite{P:2010}. As for the excited state $^3\Si^+_u$, the predicted minimum is $E_t=-2.0000409$\,Ry at $R=7.8237$\,a.u. while the accurate result is $E_t=-2.0000392$~Ry (rounded) at $R=7.85$\,a.u.~\cite{KW:1965}. It indicates to very high accuracy of the fitted curves near minima: one portion in $10^{-5} - 10^{-6}$ in energy and 4 - 3 s.d. in equilibrium distances.


\subsection{Rotational and vibrational states}

The analytic expression for the ground state potential curve $^1\Si^+_g$~\re{h2an1s} together with equation~\re{vrs} allow us to calculate the rotational and vibrational states by solving the Schr\"odinger equation. {It was done in Lagrange mesh method. The accuracy of calculated rovibrational energies is 4 -5 s.d. in comparison with accurate numerical calculations ~\cite{KW:1975} - it is certainly inside of domain of applicability of the Bohr-Oppenheimer approximation. For illustration in
Table~\ref{th21spg} some vibrational energies for two values of the angular momentum $L=0$ and $4$ are presented. The second line for each value of the vibrational quantum number $\nu$ contains the results obtained in~\cite{KW:1975}, where the adiabatic and some relativistic effects are also taken into account.}
\begin{center}
\begin{table}
\caption{Rovibrational energies $E_{\nu L}$ of the ground state $^1\Si^+_g$ for $L=0,4$
of the molecule H$_2$.
Second line are the results from~\cite{KW:1975} (rounded).}
\label{th21spg}
\begin{tabular}{rll}
\hline\hline
$\nu$&$E_{\nu0}$[Ry]&$E_{\nu4}$[Ry]\\
\hline
0  &-2.3291  & -2.3184\\
   &-2.329127& -2.318476\\
1  &-2.2911  & -2.2810\\
   &-2.291200& -2.281082\\
2  &-2.2553  & -2.2457\\
   &-2.255417& -2.245822\\
5  &-2.1605  & -2.1524\\
   &-2.160582& -2.152525\\
10 &-2.0452  & -2.0400\\
   &-2.045220& -2.040010\\
12 &-2.0165  & -2.0128\\
   &-2.016508& -2.012846\\
14 &-2.0013  &        \\
   &-2.001305&        \\
\hline\hline
\end{tabular}
\end{table}
\end{center}
In total, the $^1\Si^+_g$ potential curve supports 301 rovibrational bound states ranging from $L=0$ to $L=31$
and from $\nu=14$ to $\nu=0$, respectively, which are depicted in Figure~\ref{et2pair}
in agreement with the results of ~\cite{PK:2011}.
\begin{figure}
\begin{center}
\includegraphics[scale=1.2]{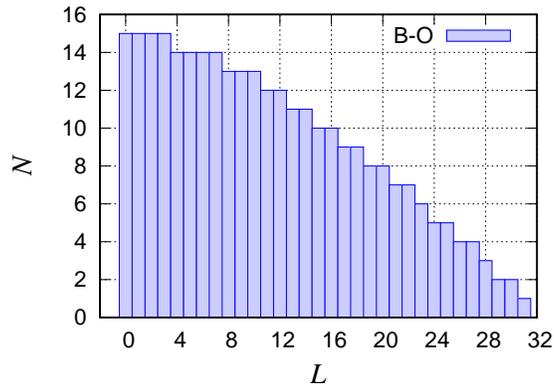}
\end{center}
\caption{Rovibrational bound states in the ground state $^3\Si_g^+$ of the molecule H$_2$ as a function of the angular momentum $L$. }
\label{et2pair}
\end{figure}
Finally, taking the analytic expression for potential curve for the excited state $^3\Si_u^+$~\re{h2an3s}, we confirm the non-existence of any vibrational or rotational bound state associated with this state in agreement with~\cite{KW:1968}. All that indicates the high quality of approximation of the potential curves for $^1\Si^+_g$ and $^3\Si_u^+$ which do not lead to any extra rovibrational bound state and do not miss any known rovibrational bound state. This feature is definitely absent in all previous attempts (known to the present authors) to build approximations of potential curves.

\section*{Conclusions}

Using accurate analytic approximations of the potential curves for $1s\si_g$ and $2p\si_u$, $2p \pi_u$ and
$3d \pi_g$ states of H$_2^+$ for the whole domain in interproton distance $R$ we study vibrational,
rotational and rovibrational states. It is shown that the ground state $1s\si_g$ can keep 420 rovibrational states within Bohr-Oppenheimer approximation which accuracy is limited to 3-4-5 figures. Going beyond the Bohr-Oppenheimer approximation to the full geometry, where finite mass effects are taken into account, the number of these states increases
surprisingly little, to 423, see \cite{M:1993}. As for vibrational states this number goes from $\nu=19$ to
$\nu=20$. At the same time as for the state $2p\si_u$ the total number of rovibrational states
(all with $\nu=0$) is equal to 3 within or beyond Bohr-Oppenheimer approximation. For the state
$2p\pi_u$ within the Bohr-Oppenheimer approximation the number of vibrational states is equal to 12
while the total number of the rovibrational bound states is equal to 284. The potential curve for
the state $3d\pi_g$ displays no minimum, thus, the state $3d\pi_g$ is pure repulsive.

The same procedure was applied to the ground state X$^2\Si^+$ of the heteronuclear system HeH.
The approximation of the potential curve gives better description of the electronic energy than
any previous approximation constructed so far.
The shallow minimum at $R\approx 6.6$\,a.u., presented by the potential curve
does not support rovibrational states. Can developed procedure of interpolation work
for other heteronuclear systems as well as homonuclear systems will be studied elsewhere.

Interestingly when the method is applied to the hydrogen molecule, a system with two electrons, the analytic expressions for the potential curves of the states $^1\Sigma_g^+$ and $^3\Sigma_u^+$ reproduce the numerical calculations in 3-4 (or more) figures. The potential curve interpolation allows us to predict accurately the position and the depth of the minimum and its position, as well as all 301 rovibrational states supported by the ground state potential curve. So far, one can only be surprised that
such a simple, straightforward interpolation provides so accurate description of both potential curves
for H$_2^+$ , HeH and H$_2$ and their rovibrational spectra with one-instanton contribution included {\it only} for homomolecular systems H$_2^+$ and H$_2$. We consider as a challenge to interpolate the lowest two potential curves for the Helium sequence:  He$_2^+$, He$_2^{2+}$ and He$_2^{3+}$~\cite{LP:1933,XPG:2005,TPA:2012} .  Even more, it is very interesting to try to approximate the potential surface for triatomic molecules. It will be done elsewhere. The results of the fits are such that all corresponding poles of the Pad\'e approximant  are complex conjugate or negative.

\bigskip

\textit{\large \bf Acknowledgements}.

\bigskip

H.O.P. is grateful to Instituto de Ciencias Nucleares, UNAM (Mexico) for a kind hospitality
extended to him where a certain stages of the present work were carried out.
The research by A.V.T. is supported in part by PAPIIT grant {\bf IN108815} (Mexico).
Present study was inspired by the question posed by F Stillinger.



\begin{thebibliography}{99}

\bibitem{LL}
        L.D.~Landau and E.M.~Lifshitz,\\
        {\it Quantum Mechanics, Non-relativistic Theory {\rm (}Course of
        Theoretical Physics {\rm vol 3)}},\\
        {3rd edn (Oxford:Pergamon Press)}, 1977

\bibitem{OT:2016}
        H.~Olivares-Pil\'on and A.V.~Turbiner,\\
        \textit{The H$_2^+$ molecular ion: Low-lying states},\\
        {\em Ann. Phys. \bf 373} (2011) 5821-608

\bibitem{Turbiner:2011}
        A.V.~Turbiner and H.~Olivares-Pil\'on,\\
        \textit{The H$_2^+$ molecular ion: a solution},\\
        {\em J. Phys. B \bf 44} 101002 (7 pp) (2011)

\bibitem{MWB:2002}
        J. N. Murrell, T. G. Wright, and S. D. Bosanac, \\
        \textit{A search for bound levels of the van der waals molecules: H$_2$(a$^3\Sigma_u^+$), HeH($X^2\Sigma^+$), LiH(a$^3\Sigma^+$) and LiHe($X^2\Sigma^+$)},\\
        {\em J. Mol. Struct.: THEOCHEM \bf 591}, 1-9 (2002)

\bibitem{BS:1966}
        W.~Byers~Brown and E.~Steiner,\\
        \textit{On the Electronic Energy of a One-Electron Diatomic Molecule near the United Atom},\\
         {\em J. Chem. Phys \bf 44}, 3934-3940 (1966)

\bibitem{K:1983}
        M.~Klaus,\\
        \textit{On H$_2^+$ for small internuclear separation},\\
        {\em J. Phys. A: Math, Gen. \bf 16}, 2709-2720 (1983)

\bibitem{BB:1965}
        W.~Byers~Brown,
        \textit{Interatomic Forces at Very Short Range},\\
        {\em Discussions Faraday Soc. \bf 40} 140-149, (1965)

\bibitem{OV:1964}
        A.A.~Ovchinkikov, and A.D.~Sukhanov,\\
        {\em Dokl.Akad.Nauk, SSSR, \bf 157}, 1092-1095 (1964),\\
        {\it Soc.Phys.-Dokl. \bf 9}, 685-687 (1965)(English translation)

\bibitem{DP:1968}
        R.J.~Damburg and R.Kh.~Propin,\\
        \textit{On asymptotic expansions of electronic terms of the molecular ion H$_2^+$},\\
        {\em J. Phys. B. (Proc Phys. Soc.) \bf  1 }, 4,  681-691 (1968)

\bibitem{Cizek:1986}
        J.~Cizek et al.,\\
        \textit{$1/R$ expansion for $H_2^+$: Calculation of exponentially small terms and asymptotics},\\
        {\em Phys. Rev. \bf A 33}, 12 - 54 (1986)

\bibitem{B:1958}
        W.A.~Bingel,\\
      \textit{United atom treatment of the behavior of potential energy curves of diatomic molecules for small R},\\
         {\em J. Chem. Phys \bf 30}, 1250-1253 (1958)

\bibitem{Baye:2015}
         D.~Baye,\\
         \textit{The Lagrange-mesh method},\\
         {\em Phys. Repts. \bf 565}, 1-107 (2015)

\bibitem{BHP:1970}
         C.L.~Beckel, B.D.~Hansen III and J.M.~Peek,\\
         \textit{Theoretical study of H$_2^+$ ground electronic state spectroscopic properties},\\
         {\em J. Chem. Phys \bf 53}, 3681-3690 (1970)

\bibitem{M:1993}
         R.~Moss,\\
         \textit{Calculations for the vibration-rotation levels of H$_2^+$ in its ground and first excited
         electronic states},\\
         {\em Mol. Phys. \bf 80}, 1541-1554 (1993)

\bibitem{P:1969}
         J.M.~Peek,\\
        \textit{Discrete Vibrational States Due Only to Long Range Forces:\ $^2\Sigma_u^+$ ($2p\sigma_u$)
          State of H$_2^+$},\\
         {\em J. Chem. Phys \bf 50}, 4595-4696 (1969)


\bibitem{BSP:1973}
         C.L.~Beckel, M.~Shafi and J.M.~Peek,\\
         \textit{Theoretical study of H$_2^+$ spectroscopic properties. II. The $2p\pi_u$ electronic state},\\
         {\em J. Chem. Phys \bf 59}, 5288-5297 (1973)


\bibitem{MF:1994}
        W.~Meyer and L.~Frommhold\\
        \textit{Long-range interactions in H-He: ab initio potential, hyperfine pressure shift and collision-induced
         absorption in the infrared},\\
        {\em Theor. Chim. Acta \bf 88}, 201-216 (1994)

\bibitem{WTT:2015}
        S.~Warnicke, K.T.~Tang, and J.P.~Toennies,\\
        \textit{Communication: Simple full range analytic potential for H$_2$, H-He, He$_2$},\\
        {\em J. Chem. Phys \bf 142} 131102 (2015)

\bibitem{Bh:1958}
         R.A.~Buckingham,\\
         {\em Trans. Faraday Soc. \bf 54}, 453 (1958)

\bibitem{YBD:1996}
         Z.C.~Yan, J.F.~Babb, and A.~Dalgarno,\\
         \textit{Variational calculations of dispersion coefficients for interactions among H, He, and Li atoms},\\
         {\em Phys. Rev. A \bf 54}, 2824-2836 (1996)

\bibitem{P:2010}
         K. Pachucki\\
         \textit{Born-Oppenheimer potential for H$_2$},\\
         {\em Phys. Rev. A \bf 82}, 032509 (2010)

\bibitem{KW:1965}
         W. Kolos and L. Wolniewicz\\
         \textit{Potential-Energy curves for the $X ^1\Sigma_g^+$, $b^3\Sigma_u^+$, and $C ^1\Pi_u$ states of the hydrogen molecule},\\
         {\em J. Chem. Phys \bf 43}, 2429 (1965)


\bibitem{HB:2001}
            M.~Hesse and D.~Baye,\\
            \textit{Lagrange-mesh calculations of excited states of three-body atoms and molecules},\\
            {\em J. Phys. B \bf 34} 1425 (2001)
\bibitem{GD:1996}
            G. W. F. Drake\\
            \textit{Atomic, Molecular and Optical Physics Handbook}  (1996)  (Springer)

\bibitem{HF:1964}
         C. Herring and M. Flicker\\
         \textit{Asymptotic exchange coupling of two hydrogen atoms},\\
         {\em Phys. Rev. \bf 134}, A362-A366 (1964)

\bibitem{KY:1967}
        I. V. Komarov and  R. K. Yanev,\\
        \textit{Molecular Term Splitting in Two-electron Exchange},\\
        {\em Soc. Phys. JETP \bf 24}, 1159 (1967)
        
\bibitem{TTY:1993}
         K. T. Tang, J. Peter Toennies, and C. L. Yiu\\
         \textit{Exchange energy of H$_2$ calculated by the surface integral method in zeroth order approximation},\\
         {\em J. Chem. Phys \bf 99}, 377 (1993)

\bibitem{BDC:2010}
         B. L. Burrows, A. Dalgarno and M. Cohen\\
         \textit{Asymptotic exchange energies for H$_2$},\\
         {\em Phys. Rev. A \bf 86}, 052525 (2012)

\bibitem{PK:2011}
         K. Pachucki and J. Komasa\\
         \textit{Gerade-ungerade mixing in the hydrogen molecule},\\
         {\em Phys. Rev. A \bf 83}, 042510 (2011)

\bibitem{BBS:1966}
         W. Byers Brown and E. Steiner\\
         \textit{On the electronic energy of a one-electron diatomic molecule near the united atom},\\
         {\em J. Chem. Phys \bf 44}, 3934-3940 (1966)

\bibitem{KW:1975}
         W. Kolos and L. Wolniewicz,\\
         \textit{Improved potential energy curve and vibrational energies for the electronic ground state of the hydrogen molecule},\\
         {\em J. Mol. Spectrosc. \bf 54}, 303-311 (1975)

\bibitem{KW:1968}
         W. Kolos and L. Wolniewicz\\
         \textit{Vibrational and Rotational Energies for the $B ^1\Sigma_u^+$, $C ^1\Pi_u$ and $a ^3\Sigma_g^+$ states of the hydrogen molecule},\\
         {\em J. Chem. Phys \bf 48}, 3672 (1968)


\bibitem{LP:1933} 
         L. Pauling,\\
         \textit{The Normal State of the Helium Molecule-Ions He$^{2+}$ and He$^{2++}$},\\
         {\em J. Chem. Phys \bf 1}, 56 (1933)

\bibitem{XPG:2005}
         J. Xie, B. Poirier, and G. I. Gellene,\\
         \textit{Accurate, two-state ab initio study of the ground and first-excited states of He$_2^+$, including exact treatment of all Born-Oppenheimer correction terms},\\
         {\em J. Chem. Phys \bf 122}, 184310 (2005)
         
\bibitem{TPA:2012}
         W.-C. Tung, M. Pavanello, and L. Adamowicz,\\
         \textit{Very accurate potential energy curve of the He$_2^+$ ion},\\
         {\em J. Chem. Phys \bf 136}, 104309 (2012)

\end{thebibliography}
\end{document}